\documentclass[amsmath,amssymb,aps,eqsecnum]{revtex4}
\usepackage[dvips]{graphicx}
 \usepackage{bm,bbm}

\begin{document}
\title{On duality between quantum maps and quantum states}
\author{Karol {\.Z}yczkowski$^{1,2}$ and Ingemar Bengtsson$^3$}

\affiliation {$^1$Instytut Fizyki im. Smoluchowskiego,
Uniwersytet Jagiello{\'n}ski,
ul. Reymonta 4, 30-059 Krak{\'o}w, Poland}
\affiliation{$^2$Centrum Fizyki Teoretycznej, Polska Akademia Nauk,
Al. Lotnik{\'o}w 32/44, 02-668 Warszawa, Poland}

\affiliation{$^3$Fysikum, Stockholm University, 
Alba Nova, S-106 91 Stockholm, Sweden}

 \date{January 20, 2004}

\begin{abstract}
          We investigate the space of quantum
          operations, as well as the larger space of maps which are positive,
          but not completely positive. A constructive criterion
          for decomposability is presented. A certain class of
          unistochastic operations, determined by 
          unitary matrices of extended dimensionality,
          is defined and analyzed. Using the concept of the
          dynamical matrix and the Jamio{\l}kowski isomorphism
          we explore the relation between the set of quantum
          operations (dynamics) and the set of density matrices
          acting on an extended Hilbert space (kinematics).
          An analogous relation is established between the classical
          maps and an extended space of the discrete probability distributions.
\end{abstract}

%
\maketitle

\medskip
\begin{center}
{\small e-mail: karol@cft.edu.pl \quad \ ingemar@physto.se}
\end{center}


\section{Introduction}

The theory of quantum information has been studied for at least forty years
-- consult  e.g. some early papers by Ingarden and his
group written in sixties and seventies \cite{IU62,IK68,Ko69,In75}.
However, rapid progress of this theory 
occurred only in the last decade \cite{Gr99,BEZ00,NC00,ABHH+01,Ke02}, 
due to a synergetic feedback with 
several recent experiments on quantum physics,
motivated by information encoding and processing.
The possibility to encode the information in quantum states
triggered an increasing interest in the structure and the properties of
the set of quantum states \cite{MaW95,AMM97,JS01,Ki03,BK03}.
Usually one considers states acting on a finite
dimensional Hilbert space, ${\cal H}_N$.

To characterize the way that information
is processed according to the laws of quantum mechanics, 
one needs to describe the dynamics
of the density matrices. In many cases it is sufficient to consider
discrete dynamics, which
maps an initial state $\rho$ into the final state $\rho'$.
Such maps are often called {\sl quantum channels},
and their theory is a subject of considerable
recent interest \cite{OP93,IKO97,AF01,BP02}.

The set of quantum states
${\cal M}^{(N)}$
 consists of density matrices of size $N$,
which are normalized, hermitian, and  positive definite.
A quantum channel is called {\sl positive},
if it maps the set of positive operators into itself.
However, since a physical system under consideration may be
coupled with an environment, the maps
describing physical processes should
be {\sl completely positive} (CP), which means that all their
extensions into higher dimensional spaces remain positive.
Any CP map may be (not uniquely) represented in the so--called Kraus
form, by a collection of Kraus operators \cite{Kr83}.

Even though positive, but not completely positive,
 maps cannot be realized in the laboratory, 
they are of a great theoretical
importance, since they can be used to detect 
entanglement in density matrices written on paper
\cite{HHH96a,Ho01}. The 
phenomenon of quantum entanglement
seems to be crucial in the theory of quantum information.
The structure of the set of all completely positive maps,
${\cal CP}_N$,
is relatively well understood \cite{Cho75a}, but
the task to characterize the larger set of all positive
maps is far from being completed \cite{TT83,Ky96,MM01}.

The aim of this work is twofold. On one hand we present a
concise review of recent development  concerning the
properties of the set of quantum maps. On the other hand,
we present several new results in this field.
In particular, in section 2 concerning the matrix algebra,
we define a simple transformation of a matrix, called {\sl reshuffling}, and
establish a useful lemma:
The Schmidt coefficients of a matrix $A$, treated as an element of a
composite Hilbert-Schmidt space of operators,
are equal to the squared singular values of the reshuffled matrix $A^R$.
In section 3 we investigate the properties of the dynamical matrix $D$
associated with any linear map \cite{SMR61} - it may be obtained by
reshuffling of the superoperator $L$ which describes the map.
Furthermore, we define a class of
{\sl unistochastic maps}, which are determined by a single unitary
matrix $U$ of size $N^2$.

For any CP map the corresponding
dynamical matrix  $D$ is hermitian and positive definite.
Making use of the eigen representation of $D$ one may define 
the canonical Kraus form of the operation.
In section 5 we present a comparison between different classes of
quantum maps, acting on the space of density matrices, with the
classes of classical maps acting on the simplex $\Delta_{N-1}$
of discrete, $N$--point probability measures.
For concreteness we present in section 6 certain exemplary maps
acting in the space of qubits - the density matrices of size $N=2$.

In section 7 the
positive, co-positive and decomposable maps are analyzed
and a constructive criterion for decomposability of a map
is provided. Furthermore, 
for any quantum map we introduce three quantities useful to locate it 
with respect to the boundaries of the sets of 
positive (CP, CcP) maps.
In the subsequent sections we explore the
Jamio{\l}kowski isomorphism \cite{Ja72}, which
relates the set of quantum maps acting on ${\cal M}^{(N)}$,
 with the states from ${\cal M}^{(N^2)}$, which act on the extended
Hilbert space ${\cal H}_N \otimes {\cal H}_N$.
This isomorphism, formulated as well in the quantum, as well as in the
classical setup, allows us to discover an
analogy between objects, sets and problems 
concerning quantum maps and quantum states. 
This key issue of the work explains its title:
there exists a kind of duality between
between properties of the set of quantum maps (dynamics)
and the the set quantum states of a
composite  $N\times N$ system (kinematics).

\section{Algebraic detour: matrix reshaping and reshuffling}

In this section we are going to discuss some
simple algebraic transformations performed on complex matrices,
which prove useful in description
of quantum maps. In particular, we introduce
a convenient notation to work in the composite 
Hilbert space ${\cal H}_N \otimes {\cal H}_M$ or
in the Hilbert--Schmidt (HS) space
of linear operators, ${\cal H}_{\rm HS}$.

Consider a rectangular matrix $A_{ij}$, $i=1,...,M$
and $j=1,...,N$. The matrix may be
 {\sl reshaped},
by putting its elements row after row
 in lexicographical
order  into a vector $\vec{a}_k$ of size $MN$,
\begin{equation}
 {\vec a}_k =A_{ij}
 \quad {\rm where} \quad
 k=(i-1)N+j, \quad i=1,...,M, \quad j=1,...N .
\label{matvec}
\end{equation}

Conversely, any vector of length $MN$ may be reshaped into a rectangular
matrix. The simplest example of such a vectorial
 notation of matrices reads
 \begin{equation}
  A \ = \   \left[
\begin{array}{cc}
A_{11} & A_{12} \\
A_{21} & A_{22}
\end{array} \right]
\quad \longleftrightarrow \quad
 {\vec a}\  = \ \{ A_{11},A_{12},  A_{21}, A_{22} \}.
\label{matvec2}
\end{equation}
The scalar product between any two elements of the HS space
${\cal H}_{\rm HS}$ (matrices of size $N$) may be rewritten as an ordinary
scalar product between two corresponding vectors of size $N^2$,
\begin{equation}
\langle A |B\rangle \equiv {\rm Tr} A^{\dagger}B = {\vec a}^* \cdot {\vec b}
= \langle a|b\rangle.
\label{Hsscalprod}
\end{equation}
Thus the HS norm of a matrix
is equal to the norm of the associated vector,
$||A||^2_{HS}=|{\vec a}|^2$.

Sometimes we will use both indices and
label a component of ${\vec a}$ by $a_{ij}$.
This vector of length $MN$ may be linearly transformed into $a'=Ca$
by a matrix $C$ of size $MN \times MN$.
Its elements may be denoted by $C_{kk'}$ with $k,k'=1,...,MN$,
but it is also convenient to use a
four index notation, $C_{\stackrel{\scriptstyle m\mu}{n \nu}}$
where $m,n=1,...,N$ while  $\mu, \nu =1,...,M$.
In this notation the elements of the transposed matrix
are $C_{\stackrel{\scriptstyle m\mu}{n\nu}}^T
=C_{\stackrel{\scriptstyle n \nu}{m \mu}}$,
since the upper pair of indices determines the row
of the matrix $C$, while the lower pair determines its column.
The matrix $C$ may represent an operator acting in
a composite space ${\cal H}={\cal H}_N \otimes {\cal H}_M$.
The tensor product of any two bases in both subspaces
provides a basis in ${\cal H}$, so that
\begin{equation}
 C_{\stackrel{\scriptstyle m \mu}{n \nu}}
= \langle e_m \otimes f_{\mu} |C|
  e_n \otimes f_{\nu} \rangle,
\label{tensbasisC}
\end{equation}
where the Roman indices refer to variables of the first subsystem,
${\cal H}_A={\cal H}_N$, and the Greek indices to the second,
${\cal H}_B={\cal H}_M$.
 The trace of a matrix reads
Tr$C=C_{\stackrel{\scriptstyle m \mu}{m \mu}}$, where summation over the
repeating indices is assumed. The operation of partial trace over the
second subsystem produces the matrix $C^A \equiv {\rm Tr}_B C$ of size $N$,
while tracing over the first subsystem leads to
a $M\times M$ matrix $C^B \equiv {\rm Tr}_A C$,
\begin{equation}
 C^A_{mn}=C_{\stackrel{\scriptstyle m \mu}{n \mu }},
{\rm \quad  \quad and \quad  \quad}
C^B_{\mu\nu}=C_{\stackrel{\scriptstyle m \mu}{m \nu}}.
\label{partraceC}
\end{equation}
If $C$ is a tensor product, $C=A\otimes B$, then
$C_{\stackrel{\scriptstyle m \mu}{n \nu}}=A_{mn}B_{\mu\nu}$.

Consider a unitary matrix $\tilde U$ of size $N^2$.
Its $N^2$ columns (rows)
${\vec{\tilde u}}_k, k=1,...,N^2$ reshaped into square matrices
${\tilde U}_k$ of size $N$ form an orthogonal basis in ${\cal H}_{HS}$. Using
the Hilbert-Schmidt norm,   $||A||_{HS}=({\rm Tr}A A^{\dagger})^{1/2}$,
we normalize them according to $A_k={\tilde U}_k/||\tilde{U}_k||_{HS}$ and
obtain the orthonormal basis,   $\langle A_k|A_j\rangle \equiv {\rm Tr}
A_k^{\dagger}A_j = \delta_{kj}$.  Alternatively, in a double index notation
with $k=(m-1)N+\mu$ and $j=(n-1)N+\nu$
this orthogonality relation reads
$\langle A^{m\mu}|A^{n\nu}\rangle= \delta_{mn}\delta_{\mu\nu}$.
Note that in general the matrices $A_k$
(also denoted by $A^{m\mu}$)  are not unitary.

Let $X$ denote an arbitrary matrix of size $N^2$. It may be represented
as a double (quadruple) sum,
\begin{equation}
|X\rangle =  \sum_{k=1}^{N^2} \sum_{j=1}^{N^2} C_{kj}
 |A_k\rangle \otimes |A_j\rangle =
C_{\stackrel{\scriptstyle m \mu}{n \nu}}
  |A^{m\mu}\rangle \otimes |A^{n\nu}\rangle
\label{Vmatrix}
\end{equation}
where $C_{kj}={\rm Tr}( (A_k \otimes A_j)^{\dagger} X)$
may be neither Hermitian nor normal (which means that $C$ and
$C^{\dagger}$ need not commute).
The matrix $X$ may be considered as a vector in the
composite Hilbert-Schmidt space,
${\cal H}_{\rm HS}\otimes {\cal H}_{\rm HS}$,
so applying its Schmidt decomposition \cite{Pe95b}
we arrive at
\begin{equation}
|X\rangle = \sum_{k=1}^{N^2} {\sqrt{ \lambda_{k}}} |A_k^{\prime}\rangle
\otimes |A_k^{\prime\prime}\rangle,
 \label{VSchmidt}
\end{equation}
where $\sqrt{\lambda_k}$ are the singular values of $C$,
i.e. the square roots of the non-negative eigenvalues
of $CC^{\dagger}$.
The sum of their squares is determined by the norm of the operator,
 $\sum_{k=1}^{N^2} \lambda_k= {\rm Tr}(XX^{\dagger})=||X||_{HS}^2$.

Since the Schmidt coefficients  {\sl do not}
depend on the initial basis
let us analyze the special case, in which the
basis in ${\cal H}_{HS}$ is generated by the identity matrix,
$U={\mathbbm 1}$ of size $N^2$.
Then each of the $N^2$ basis matrices of size $N$
consist of only one non-zero element which equals
unity, $A_{k}=A^{m\mu}=|m\rangle\langle \mu |$, where
{\mbox{$k=N(m-1)+\mu$}}.  Their tensor products form an orthonormal basis
in  ${\cal H}_{HS}\otimes {\cal H}_{HS}$ and allow to represent an
arbitrary matrix $X$ in the form (\ref{Vmatrix}).
In this case the matrix of the coefficients $C$ has a
particularly simple form,
{\mbox{$C_{\stackrel{ \scriptstyle m \mu }{n \nu}}=
{\rm Tr}[(A^{m\mu}\otimes A^{n\nu})X]=
X_{\stackrel{ \scriptstyle mn}{\mu\nu}}$}}.

This particular reordering of a matrix deserves a name
so we shall write $X^R \equiv C(X)$ defining the following
procedure of {\sl reshuffling} of matrices.
Using this notion our findings may be summarized in the following lemma:
\smallskip

{\sl Schmidt coefficients of an operator $X$
acting on a bi-partite Hilbert space are equal to the squared
singular values of the reshuffled matrix, $X^R$.}
\smallskip

More precisely, the  Schmidt decomposition (\ref{VSchmidt}) of any operator $X$
of size $MN$ may be supplemented by a set of three equations
\begin{equation}
\left\{
\begin{array}{ccc}
 \{ \lambda_{k} \}_{k=1}^{N^2} & = &  \bigl\{ {\rm SV} (X^R) \bigr\}^2
  \ \ : \ \  { \rm \ \ eigenvalues \ \ of \ \ } (X^R)^{\dagger}X^R \\
|A^{\prime}\rangle & \ : \ &  {\rm \ \ reshaped \ \ eigenvectors \ \ of \ \ }
(X^R)^{\dagger}X^R   \\
|A^{\prime\prime}\rangle  & \ : \  &
{\rm \ \ reshaped \ \ eigenvectors \ \ of \ \ } X^R(X^R)^{\dagger} \\
\end{array} \right. ,
\label{VSchmidt2}
\end{equation}
where we have assumed that $N \le M$.
The initial basis is
transformed by a local unitary transformation $W_a \otimes W_b$,
where $W_a$ and $W_b$ are matrices of eigenvectors of matrices
$(X^R)^{\dagger}X^R$ and $X^R(X^R)^{\dagger}$, respectively.
Iff rank $r$ of $X^R(X^R)^{\dagger}$  equals one,
the operator can be factorized into a product form,
$X=X_1\otimes X_2$,  where
$X_1={\rm Tr}_2 X$ and  $X_2={\rm Tr}_1 X$.

\medskip
\begin{table}
\caption{Reorderings of a matrix $X$ representing an operator
which acts on a composed Hilbert space.
 The arrows denote the indices exchanged. }
\smallskip
\hskip 0.4cm
{\renewcommand{\arraystretch}{1.7}
\begin{tabular}
 [c]{||l|c|c|c|c||}
\hline \hline
Transformation  & definition  & symbol &
\parbox {2.4cm}{\centering preserves  \\  Hermicity  } &
\parbox {2.1cm}{\centering preserves  \\  spectrum } \\
\hline
transposition &
$X_{\stackrel{ \scriptstyle m \mu }{n \nu}}^{T}
=X_{\stackrel{ \scriptstyle n \nu}{m \mu}}$
& $\updownarrow \updownarrow $
& yes & yes \\
\hline
flip &
$X_{\stackrel{ \scriptstyle m \mu }{n \nu}}^{F}
=X_{\stackrel{ \scriptstyle \mu m}{\nu n}}$
& $  {\stackrel{   \textstyle
\leftrightarrow }{\leftrightarrow }} $
  & yes & yes \\
\hline
partial &
$X_{\stackrel{ \scriptstyle m \mu }{n \nu}}^{T_A}
=X_{\stackrel{ \scriptstyle n \mu}{m \nu}}$
& $\updownarrow . $
& yes & no \\
transpositions &
$X_{\stackrel{ \scriptstyle m \mu }{n \nu}}^{T_B}
=X_{\stackrel{ \scriptstyle m \nu}{n \mu}}$
& $. \updownarrow $
& yes & no  \\
\hline
reshuffling &
$X_{\stackrel{ \scriptstyle m \mu }{n \nu}}^{R}
=X_{\stackrel{ \scriptstyle
m n}{\mu \nu}} $
& $\nearrow \! \! \! \! \! \! \swarrow $
& no  & no \\
reshuffling $'$ &
$X_{\stackrel{ \scriptstyle m \mu }{n \nu}}^{R'}
=X_{\stackrel{ \scriptstyle
\nu \mu}{n m}}$
& $\nwarrow \! \! \! \! \! \! \searrow $
& no  & no \\
\hline
partial &
$X_{\stackrel{ \scriptstyle m \mu }{n \nu}}^{F_1}
=X_{\stackrel{ \scriptstyle \mu m}{n \nu}}$
& $  {\stackrel{   \textstyle
\leftrightarrow }{ . }} $
& no & no \\
flips &
$X_{\stackrel{ \scriptstyle m \mu }{n \nu}}^{F_2}
=X_{\stackrel{ \scriptstyle m \mu}{\nu n}}$
& $  {\stackrel{   \textstyle  . }
{\leftrightarrow }} $
& no & no  \\
\hline \hline
\end{tabular}
}
\label{tab:Xmnmunu}
\vskip -0.55cm
\end{table}
\medskip

In general,
one may reshuffle square matrices, if its size $K$ is not prime.
The symbol $X^R$ has a unique meaning if a concrete decomposition
of the size $K=MN$ is specified. If $M\ne N$ the matrix
$X^R$ is a $N^2 \times M^2$ rectangular matrix. Since
$(X^R)^R=X$ we see that one may also reshuffle rectangular matrices,
provided both dimensions are squares of natural numbers.
Similar reorderings of matrices
were considered by Hill et al. \cite{OH85,YH00}
while investigating CP maps 
and later in \cite{Rud02,HHH02,CW03,Rud03,ACF03}
to analyze separability of mixed quantum states.

To get a better feeling of the reshuffling transformation
observe that reshaping each row of an initially square matrix $X$ of size $MN$
according to Eq. (\ref{matvec}) into a
rectangular $M \times N$ submatrix, and
placing it  according to the lexicographical order block after block,
one produces the reshuffled matrix $X^R$.
Let us illustrate this procedure for the simplest case $N=M=2$,
in which any row of the matrix $X$ is reshaped into a $2 \times 2$ matrix
\begin{equation}
C_{kj}=X_{kj}^R  \equiv \left[
\begin{array}{c|c}
{\bf {X_{11}\ \ X_{12}}}  & X_{21}{\rm ~ ~ ~ }X_{22} \\
X_{13} {\rm ~ ~ ~ }X_{14} & {\bf X_{23} \ \ X_{24}} \\
\hline
{\bf X_{31}\ \ X_{32}} & X_{41} {\rm ~ ~ ~ }X_{42} \\
X_{33}{\rm ~ ~ ~ }X_{34} & {\bf X_{43}\ \ X_{44} }
\end{array}
\right] .
\label{reshuf1}
\end{equation}
The operation of reshuffling could be defined in an alternative way, say
the reshaping of the matrix $A$ from (\ref{matvec}) could be performed
column after column into a
vector ${\vec a}^{\prime}$.
In the four indices notation introduced above
(Roman indices running from $1$ to $N$ correspond to the first
subsystem, Greek indices to the second one),
both operations of reshuffling take the form
  \begin{equation}
X_{\stackrel{ \scriptstyle m \mu }{n \nu}}^{R}
\equiv X_{\stackrel{
\scriptstyle
m n}{\mu \nu}}
{\quad \quad  \quad \rm and \quad \quad \quad}
X_{\stackrel{ \scriptstyle m\mu}{ n \nu}}^{R'} \equiv
X_{\stackrel{ \scriptstyle
\nu \mu}{n m}}  .
 \label{reshuff}
 \end{equation}

However, both reshuffled matrices are equivalent up to
a certain permutation of rows and columns and transposition, so
the singular values of  $X^{R^{\prime}}$ and $X^R$ are equal.
It is
easy to see that $(X^{R})^{R}=X.$ In the symmetric case with $M=N$,
 $N^{3}$ elements of $X$
do not change their position during the operation of reshuffling
(these are typeset {\bf boldface} in (\ref{reshuf1})); while the other
$N^{4}-N^{3}$ elements are exchanged. The space of complex matrices
with the reshuffling symmetry, $X=X^R$,
is thus $2N^4-2(N^4-N^3)=2N^3$ dimensional.

For comparison we provide analogous formulae showing the action of
{\sl partial transposition}: with respect to the first subsystem,
$T_A\equiv T\otimes {\mathbbm 1}$ and with respect to the second,
$T_B \equiv {\mathbbm 1} \otimes T$,
\begin{equation}
X_{\stackrel{ \scriptstyle m \mu}{n\nu}}^{T_A}=
X_{\stackrel{ \scriptstyle n\mu}{m\nu}}
 {\quad \quad  \rm and \quad \quad}
X_{\stackrel{ \scriptstyle m \mu}{n \nu}}^{T_B}
=X_{\stackrel{ \scriptstyle m \nu }{n \mu}}.
\label{partaltra}
 \end{equation}
Note that all these operations consist of exchanging a given pair of
 indices.
However, while partial transposition (\ref{partaltra})
preserves Hermicity, the reshuffling (\ref{reshuff}) does not.
For convenience we shall define a related transformation
of {\sl flip} among both subsystems,
$X_{\stackrel{ \scriptstyle m \mu }{n \nu}}^{F} \equiv
X_{\stackrel{ \scriptstyle \mu m}{\nu n }}$,
the action of which consists in relabeling of
certain rows (and columns) of the matrix, so
its  spectrum remains preserved.
Note that for a tensor product $X=Y\otimes Z$ one has $X^F=Z\otimes Y$.
In full analogy to partial transposition we use also
two operations of {\sl partial flip}
(see table \ref{tab:Xmnmunu}).
All the above transformations are involutions, since performed
twice they are equal to identity.
It is not difficult to find relations between them, e.g.
 $X^{F_1}=[(X^{R'})^{T_A}]^{R'}=[(X^R)^{T_B}]^R$.
Since  $X^{R'}=[(X^R)^F]^T=[(X^R)^T]^F$, while
$X^{T_B}=(X^{T_A})^T$
and $X^{F_1}=(X^{F_2})^F$,
thus the spectra and singular values of the reshuffled
(partially transposed, partially flipped) matrices
do not depend on the way, each operation has been performed,
i.e.
 eig$(X^R)={\rm eig}(X^{R'})$ and  SV$(X^R)={\rm SV}(X^{R'})$,
 (eig$(X^{F_1})={\rm eig}(X^{F_2})$ and SV$(X^{F_1})={\rm SV}(X^{F_2})$).
\medskip

\section{Completely positive maps}
\label{sec:cpmap}

Let $\rho\in {\cal  M}^{(N)}$ be a density operator acting
on a $N$-dimensional Hilbert space ${\cal H}_N$.
What conditions need to be fulfilled by a
map $\Phi:{\cal  M}^{(N)}\to {\cal  M}^{(N)}$,
so it could represent a physical operation?
If the map is {\sl linear} than the image of a mixed state $\rho$
does not depend on the way, how the state $\rho$ was constructed out
of projectors. This is a very important feature, since it
allows for a probabilistic interpretation of any mixed state.
Furthermore, we assume that the map is trace preserving,
Tr$\rho={\rm Tr}\Phi(\rho)$,
which corresponds to the conservation of probability.

Any quantum operation has to be
 {\sl positive} -- it should map any positive density
operator into a positive operator. Strictly speaking the term
'positive' refers to {\sl positive semidefinite} Hermitian
operators, which do not have negative eigenvalues. However, this condition
occurs not to be sufficient to produce physically realizable
transformations.  Any quantum state $\rho$ may be extended
by an ancilla  $\sigma$ into a
tensor product acting in a
$KN$ dimensional Hilbert space.
Hence the evolution of a linear
map $\Phi$ may be considered as
the evolution of a part $\rho$ of a larger system $\rho\otimes \sigma$.
Therefore we should require {\sl complete positivity} \cite{St55,Ar69},
which means that for an arbitrary $K$ dimensional extension
\begin{equation}
 {\cal H}_N \to {\cal H}_N \otimes {\cal H}_K
\quad {\rm  the\quad  map} \quad
\Phi\otimes {\mathbbm 1}_K
\quad {\rm is \quad positive}.
 \label{cpamap}
\end{equation}
If the above requirement holds for any fixed $K$, such a property
is called $K$--{\sl positivity} \cite{St55,Da76,Kr83}.
Constructing the dynamical matrix we will show in the next section
that any $N$--positive map $\Phi$
acting on ${\cal  M}^{(N)}$ is completely positive \cite{Cho72}.
The importance of complete positivity 
in quantum mechanics was emphasized in the seventies by 
Kraus \cite{Kr71}, Lindblad \cite{Li76}  and Accardi \cite{Ac76},
and the restriction to CP maps is not trivial:
There exist quantum maps which are positive but not completely
positive \cite{St55}.
A simple and important example consists of
transposition $T$ of an operator in a given basis \cite{Per96},
which for Hermitian operators is equivalent to complex conjugation:
 if $A=A^{\dagger}$ then $A^{T}={\bar A}$.
Although the map $T(\rho)= \rho^T$ preserves the spectrum and
therefore is positive, its extension  $T_A \equiv  T \otimes {\mathbbm 1}$,
called {\sl partial transposition}, is not, as
discussed in section \ref{sec:positiv}.

A linear CP-map which preserves the
trace is called a {\sl quantum operation}
or {\sl quantum channel}.
Any linear, completely positive map $\Phi$ may be represented
\cite{Kr71,Kr83}
by a collection of $k$ {\sl Kraus operators} $A_i$
in the so--called
{\sl Kraus form}
\begin{equation}
\rho\to \rho'={\Phi}(\rho)=
\sum_{i=1}^k A_i\rho A_i^{\dagger} .
 \label{Kraus1}
\end{equation}
It is also called an {\sl operator sum representation} of the map $\Phi$
or the {\sl Stinespring form},
since its existence follows from the Stinespring dilation theorem
 \cite{St55,Ev84}. If the set of Kraus
operators satisfies the completeness relation
\begin{equation}
 \sum_{i=1}^k A_i^{\dagger} A_i={\mathbbm 1} ,
 \label{Kraus2}
\end{equation}
the map is trace preserving:
 for any initial state $\rho$ one has
 ${\rm Tr} \bigl( \sum_{i=1}^k A_i^{\dagger} A_i\rho \bigr)=$
 ${\rm Tr}\bigl[ {\Phi(\rho)} \bigr] = {\rm Tr }\rho=1$.
Thus, the Kraus operators may be considered as
measurement operators while any trace preserving CP map may be interpreted as a
generalized measurement 
\cite{Pe95b}.

Let us denote the elements of the Kraus operators $A_i$
represented in an orthonormal basis by $A^{(i)}_{mn}$.
Define a {\sl Kraus matrix} $M$ of size $N$
composed of  non-negative entries \cite{St02},
\begin{equation}
  M_{mn}= \sum_{i=1}^k |A_{mn}^{(i)} |^2,
 \label{KrausMM}
\end{equation}
where the Kraus operators are now denoted by $A^{(i)}$.
The completeness relation (\ref{Kraus2})
enforces that the sum of elements in each column of $M$ equals to unity,
so it is a stochastic matrix.
Thus quantum operations are often called  {\sl stochastic maps}.

\vskip 0.1cm
\begin{figure} [htbp]
   \begin{center}
\
 \includegraphics[width=12.3cm,angle=0]{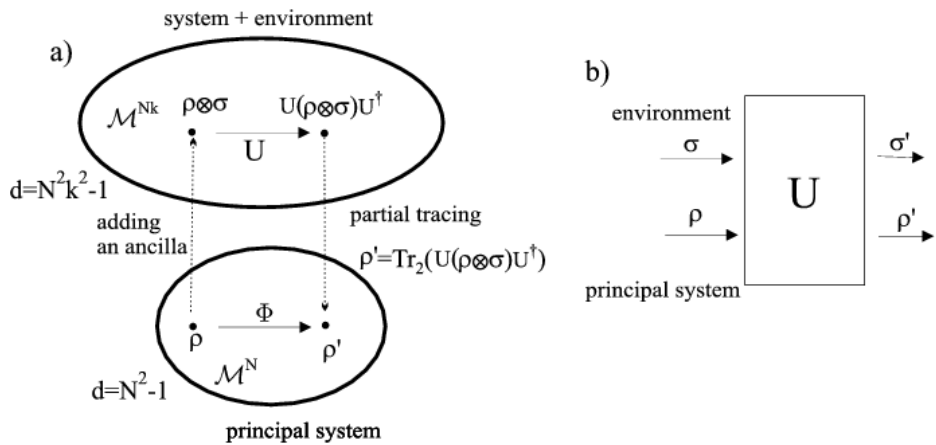}
\caption{
 Quantum operations represented by
a) unitary operator $U$ of size $NK$
 in an enlarged system including the
 environment, b) black box picture, in which $U$ couples
 the principal $N$--dimensional system $\rho$ with the environment
  $\sigma$ of dimensionality $K$.}
 \label{fig:oper2}
\end{center}
 \end{figure}

Any map written in the Kraus form
(\ref{Kraus1}) is linear and preserves the trace if
the condition (\ref{Kraus2}) is fulfilled.
Let $\rho$ denote the density operator of
the investigated quantum system.
The time evolution of an isolated quantum system is
unitary, while all sorts of a non unitary dynamics
reflect its interaction with an environment, described by
a density operator $\sigma$.
However, if we extend the system and study the dynamics of the total system,
$\rho \otimes \sigma$,
composed of the system under investigation and the environment,
its time evolution remains unitary \cite{Ar69,EL77}.
The non unitary
operation of $\rho$ emerges as an effect of
the partial tracing with respect to the environment,
\begin{equation}
\rho\to \rho'={\Phi}(\rho)=
{\rm Tr}_{\rm env} \Bigl[ U \bigl( \rho\otimes \sigma \bigr)
 U^{\dagger} \Bigr] .
\label{uniteform}
\end{equation}
The above form is called as {\sl environmental representation}
of the map $\Phi$.
The entire process may be considered as a composition of
three basic steps:  adding an ancilla,
 unitary transformation,
 and  partial tracing.
The forms (\ref{Kraus1}) and (\ref{uniteform})
are equivalent in the sense that any quantum operation may be
written either way.
Both representations are not unique, but as
discussed in section \ref{sec:dynmat},
the notion of a dynamical matrix
allows one to define a distinguished, canonical Kraus form.


Let the principal $N$--dimensional system $\rho$
be subjected to an operation defined by the $kN$ dimensional
matrix $U$ and an initially pure state of the
 environment
 $\sigma=|\nu\rangle  \langle \nu|$,
\begin{equation}
\rho'={\Phi}(\rho)=
{\rm Tr}_{\rm env} \Bigl[ U \bigr( \rho\otimes |\nu\rangle \langle
\nu |\bigl)   U^{\dagger} \Bigr]
= \sum_{\mu =1}^k  \langle \mu |U| \nu \rangle
\rho    \langle \nu |U^{\dagger}| \mu \rangle .
  \label{unitevol2}
\end{equation}

Since the fixed environmental  state $|\nu \rangle$
belongs to the $k$-dimensional Hilbert space ${\cal H}_{\rm env}$,
the expression $\langle \mu |U| \nu \rangle$ represents a square matrix
of size $N$, which we shall call $A_{\mu}$.
In the double index notation, introduced above, any element of the
unitary matrix is denoted by $U_{\stackrel{ \scriptstyle m\mu}{n \nu}}$,
while the matrix $A_{\mu}=\langle \mu |U| \nu \rangle$
consists of elements

\begin{equation}
A^{(\mu)}_{mn}= \langle m, \mu |U|n, \nu \rangle=
U_{\stackrel{\scriptstyle m \mu}{n \nu }}.
  \label{matunitrunc}
\end{equation}

 It is easy to see that  (\ref{unitevol2})
takes the operator sum form, $\sum_{\mu=1}^k A_{\mu} \rho A_{\mu}^{\dagger}$.
Moreover, due to unitarity of $U$ the operators $A_{\mu}$
satisfy the completeness relation
  \begin{equation}
\sum_{\mu =1}^k A_{\mu} ^{\dagger} A_{\mu} =
\sum_{\mu =1}^k  \langle \nu |U^{\dagger}|\mu \rangle \langle \mu |
U| \nu \rangle  = \langle \nu | U^{\dagger} U| \nu \rangle ={\mathbbm 1}_N
  \label{unitevol3}
\end{equation}
and may be considered as Kraus operators.
Every element of a given matrix $A_{\mu}$
belongs to one of the $N^2$ blocks of size $k$ of the
unitary matrix $U$. However, if we reorder it constructing
 $U_{\stackrel{ \scriptstyle m \mu }{n \nu}}^{F}
 =U_{\stackrel{ \scriptstyle \mu m}{\nu n}}$,
then each matrix $A_{\mu}$ is just a
truncation of the flipped matrix $U^F$,
 since  $A^{(\mu)}_{mn}=U_{\stackrel{ \scriptstyle \mu m}{\nu n}}^F$
represents its minor.

If the initial state of the
environment in the representation (\ref{uniteform})
is chosen to have full rank,
$\sigma=\sum_{\nu=1}^k q_{\nu} |\nu \rangle \langle \nu|$,
 the operator sum representation consists of $kN$ terms
\begin{equation}
\rho'={\Phi}(\rho)=
{\rm Tr}_{\rm env} \Bigl[ U \Bigr( \rho\otimes \sum_{\nu=1}^k q_{\nu}
|\nu \rangle
 \langle \nu|\Bigl)   U^{\dagger} \Bigr]
= \sum_{l=1}^{kN} A_l \rho A_l^{\dagger}
 \label{unitevol4}
\end{equation}
where $A_l= \sqrt{q_{\mu}} \langle \mu |U| \nu\rangle $
and $l=\mu+\nu(k-1)$.

In this way we have shown that for any operation
written in the environmental form (\ref{uniteform})
we may find a corresponding Kraus form. Conversely,
for any quantum operation in the Kraus form
(\ref{Kraus1}) consisting of $k$ operators
we may construct an environmental form \cite{Ar69,Li75}.
For instance take the first $N$ columns of a unitary matrix $U$ of
size $kN$ are constructed of $k$ Kraus operators as defined in
(\ref{matunitrunc}). Due to the completeness relation (\ref{Kraus2})
they are normalized and orthogonal. The matrix $U$
has to be completed by $N(k-1)$ complex orthogonal vectors;
they do not influence the quantum operation and may be selected
arbitrarily.

If the initial state of the environment is pure
its dimensionality $k$ of ${\cal H}_{\rm env}$
needs not to exceed $N^2$,
the maximal number of Kraus operators required.
If the environment is initially in a mixed state,
its weights coefficients $q_j$ are needed to specify the operation.
Counting the number of parameters one could thus speculate
that the action of any quantum operation
may be simulated by a coupling with a mixed state of the environment
of size $N$. However, this is not the case: already for $N=2$
there exist operations which have to be simulated
with $3$--dimensional environment \cite{TCDGS99,ZR02}, and the
general question of the minimal size of  ${\cal H}_{\rm env}$
remains open.

It is also illuminating to discuss a special case of the
problem,
in which the initial state of the $N$--dimensional
 environment is maximally mixed,
$\sigma=\rho_*={\mathbbm 1}_N/N$.
The unitary matrix $U$ of size $N^2$, defining the
map, may be treated as a vector in the composed Hilbert--Schmidt
space ${\cal H}_{\rm HS} \otimes {\cal H}_{\rm HS}$
and represented in its Schmidt form (\ref{VSchmidt}),
$U=  \sum_{i=1}^{N^2} \sqrt{\lambda_i}  |{\tilde A}_i\rangle \otimes
 | {\tilde A}_i'\rangle $, where $\lambda_i$ are
eigenvalues of $(U^R)^{\dagger}U^R$.
Since the operators ${\tilde A}_i'$, obtained by reshaping eigenvectors
of $(U^R)^{\dagger}U^R$, form an orthonormal basis in
${\cal H}_{\rm HS}$, the procedure of partial tracing leads
to the Kraus form consisting of $N^2$ terms:

\begin{eqnarray}
\rho'  & ={\Phi}_U(\rho)=  {\rm Tr}_{\rm env} \Bigl[U(\rho\otimes
\frac{1}{N} {\mathbbm 1}_N) U^{\dagger} \Bigl] \quad\quad\quad \quad \quad  &
\nonumber
\\
 & {\displaystyle = {\rm Tr}_{\rm env} \Bigl[ \sum_{i=1}^{N^2}
\sum_{j=1}^{N^2} \sqrt{\lambda_i\lambda_j}
\bigl( {\tilde A}_i \rho {\tilde A}_j^{\dagger} \bigr) \otimes
\bigl( \frac{1}{N} {\tilde A}_i' {{{\tilde A}_j}}^{'\dagger } \bigr)
\Bigr] }
 & = \frac{1}{N} \sum_{i=1}^{N^2} \lambda_i {\tilde A}_i \rho
{\tilde A}_i^{\dagger} .  
\label{unitevol5}
\end{eqnarray}
Operations, for which there exist a unitary matrix $U$
providing a representation in the above
form, we shall call {\sl unistochastic channels}.
In a way, these
 maps are analogous to classical transformations given by
unistochastic matrices, ${\vec p}'=T{\vec p}$, where
 $T_{ij}=|U_{ij}|^2$, since
both dynamics are uniquely determined by a unitary matrix
$U$ -- of sizes $N$ and $N^2$ in the classical
and the quantum cases, respectively.

In general one may consider maps analogous to (\ref{unitevol5})
with an arbitrary size of the environment. In particular
we define generalized, {\sl $K$--unistochastic maps}.
determined by a unitary matrix $U(N^{1+K})$,
in which the environment of size $N^K$
is initially in the maximally mixed state,
$N^{-K} {\mathbbm 1}_{N^K}$.
Such operations were analyzed
in context of quantum information processing \cite{KL98,PBLO03},
and, under the name 'noisy maps', by studying reversible transformations from
pure to mixed states \cite{HHO03}.
By definition, $1$--unistochastic maps are unistochastic.

For any unistochastic map
the standard Kraus form  (\ref{Kraus1})
is obtained by rescaling
the operators, $A_i =\sqrt{\lambda_i/N}\ {\tilde A}_i$.
Note that the matrix $U$ is determined up to a local unitary
matrix $V$ of size $N$, in sense that $U$ and $U'=U({\mathbbm 1} \otimes
V)$ generate the same unistochastic map, $\Phi_U=\Phi_{U'}$.

\section{Dynamical matrix}
\label{sec:dynmat}

A density matrix $\rho$ of finite size $N$ may be treated as
a vector $\vec \rho$ reshaped according to (\ref{matvec}).
The action of a linear superoperator $\Phi:\rho\to \rho'$ may thus be 
represented by a matrix $L$ ("L" like linear) of size $N^2$
\begin{equation}
{\vec \rho}\ '= L {\vec \rho}
\quad \quad {\rm or} \quad \quad
\rho_{m\mu}\!\!\!\!\!\!'\ \ = L_{\stackrel{\scriptstyle m \mu}{n \nu}}
  \rho_{n \nu} ,
 \label{dynmatr1}
\end{equation}
where summation over repeated indices is understood.
Nonhomogeneous linear maps ${\vec \rho}\ '= L' {\vec \rho} +
\vec {\sigma}$ may also be described in this way.
To obtain the homogeneous form (\ref{dynmatr1})
it suffices to substitute the matrix
$L_{\stackrel{\scriptstyle m \mu}{n \nu}}'$ by
$L_{\stackrel{\scriptstyle m \mu}{n \nu}}=
L_{\stackrel{\scriptstyle m \mu}{n \nu}}'+\sigma_{m \mu}\delta_{n\nu}$.

We require that the image $\rho'$ is a density matrix,
so it is Hermitian, positive, and normalized. These
three conditions impose constraints on the matrix $L$:
\begin{eqnarray}
{\rm \bf  i)  \quad }    & \quad
\rho'= (\rho')^{\dagger} &
 \quad \Longrightarrow \quad
L_{\stackrel{\scriptstyle \mu m}{\nu n}} =
L^*_{\stackrel{\scriptstyle m\mu}{n \nu}}
{\quad \rm so \quad}
L^*=L^F,
 \label{dynmatr2a}         \\
{\rm \bf ii)  \quad} & \quad
\rho' \ge 0          &
 \quad \Longrightarrow \quad
L_{\stackrel{\scriptstyle m\mu }{n\nu}} \rho_{ n \nu} \ge 0
{\rm \quad for \quad any \quad  state \quad }
\rho,
 \label{dynmatr2b}
\\
{\rm \bf iii)  \quad} & \quad
{\rm tr} \rho'= 1  &
 \quad \Longrightarrow \quad
\sum_{m=1}^N L_{\stackrel{\scriptstyle mm}{\mu \nu}} =
\delta_{\mu\nu} .
\label{dynmatr2c}
\end{eqnarray}
Note that (\ref{dynmatr2a}) is not the condition
of Hermicity and in general the matrix $L$ representing the operation $\Phi$
is not Hermitian.
However, if we reshuffle it according to  (\ref{reshuff})
and define the {\sl dynamical matrix}
\begin{equation}
D_{\Phi} \equiv L^R
 \quad {\rm so \quad that }\quad
D_{\stackrel{\scriptstyle m \mu}{n \nu}} =
L_{\stackrel{\scriptstyle m n }{\mu \nu}},
 \label{dynmatr3}
\end{equation}
than $D_{\Phi}$ is Hermitian, $D=D^{\dagger}$,
due to (\ref{dynmatr2a}).
The linear superoperator $L$ is then uniquely determined
by the dynamical matrix, since $L=D^R$.
The notion of dynamical matrix was introduced
already in 1961 by Sudarshan, Mathews and Rau \cite{SMR61}.
Later such matrices were used by Choi \cite{Cho75a} and are sometimes
called {\sl Choi matrices} \cite{Ha03}.

What conditions must be satisfied by a Hermitian matrix $D$
of size $N^2$ to be a dynamical matrix?
The trace condition (\ref{dynmatr2c}),
rewritten  below for the matrix $D$,
determines its trace
\begin{equation}
{\rm Tr}_A D =
\sum_{m=1}^N D_{\stackrel{\scriptstyle m\mu}{m\nu}} =
\delta_{\mu\nu}
\quad \Longrightarrow \quad
{\rm Tr}D =
\sum_{m=1}^N \sum_{\mu=1}^N
D_{\stackrel{\scriptstyle m\mu}{m\mu}} =
\sum_{\mu=1}^N \delta_{\mu \mu}=N.
\label{dynmatr3c}
\end{equation}
The positivity condition (\ref{dynmatr2b})
implies that for any states $|x\rangle$ and $\rho$
the expectation value $\langle x |D^R \rho|x\rangle$
is not negative.
Assuming that the initial state is pure,
$\rho=|y\rangle\langle y|$  so that
$\rho_{\mu\nu}=y_{\mu}y^*_{\nu}$, we obtain
 \begin{equation}
\langle x| D^R (|y\rangle \langle y|) |x \rangle =
\bigl(\langle x|\otimes \langle y|\bigr)  D
\bigl( | y\rangle \otimes |x \rangle \bigr) \ge 0,
 \label{dynmatr3b}
\end{equation}
since the double sum
 $D_{\stackrel{\scriptstyle m \mu}{n \nu}}y_{\mu} y_{\nu}^*$
represents as well the matrix $D^R (|y\rangle \langle y|)$
as well as $\langle y|  D | y\rangle$.
Thus the dynamical matrix $D$ must be positive on
product states $|y\rangle \otimes |x\rangle$.
This property is called {\sl block--positivity}.
In fact Jamio{\l}\-kow\-ski proved, in 1972, that
the converse is also true:
if property (\ref{dynmatr3b}) is satisfied then the map
$\Phi$ determined by matrix $D$ is positive \cite{Ja72} --
see section \ref{sec:jamiol}.

Positivity of $D_{\Phi}$ is a sufficient, albeit
not necessary requirement for (\ref{dynmatr3b}).
If $D_{\Phi}\ge 0$ then the map $\Phi$ is completely positive
\cite{Cho75a,PH81,FA99,Ha03,SS03}.
To prove this let us represent $D_{\Phi}$ by its spectral decomposition,
 \begin{equation}
D_{\Phi}=\sum_{i=1}^{k} d_i |\chi_i\rangle \langle \chi_i|
\quad {\rm so \quad that \quad}
D_{\stackrel{\scriptstyle m\mu}{n\nu}} =
\sum_{i=1}^{k} d_i
\chi^{(i)}_{m\mu} {\bar \chi}^{(i)}_{n\nu} \ .
\label{dynmatr4}
\end{equation}
Here $k$ denotes the rank of $D_{\Phi}$ so $k\le N^2$, while
 $\chi^{(i)}_{m\mu}$ represent $N \times N$ matrices
obtained by reshaping the eigenvectors $|\chi_i\rangle$
of length $N^2$.
Due to (\ref{dynmatr3c}) the sum of all eigenvalues $d_i$
is equal to $N$. If $D_{\Phi}$ is positive,
all its eigenvalues
are nonnegative, so we may rescale the eigenvectors
defining the operators $A_i=A^{(i)}$
\begin{equation}
 A^{(i)}_{m\mu} \equiv  \sqrt{d_i} \chi^{(i)}_{m\mu}, \quad i=1,2,...,k,
 \label{dynmatr5}
\end{equation}
so that
\begin{equation}
D_{\stackrel{\scriptstyle m\mu}{n\nu}} =
\sum_{i=1}^{k}
A^{(i)}_{m\mu} {\bar A}^{(i)}_{n\nu}
=\sum_{i=1}^{k}
  \bigl( A_i \otimes {\bar A}_i \bigr)^R .
\label{dynmatr5b}
\end{equation}
If all Kraus operators are real
$A_i={\bar A}_i$, or purely imaginary, $A_i=-{\bar A}_i$,
then the Hermitian dynamical matrix is real and hence symmetric,
 $D_{\Phi}=D_{\Phi}^T$.
Reshuffling the dynamical matrix $D$
(of the size $N^2$, with a finite $N$), we may write
\begin{equation}
L=\sum_{i=1}^{k}
  A_i \otimes {\bar A}_i
=\sum_{i=1}^{k}
   d_i \chi^{(i)} \otimes {\bar \chi}^{(i)}.
\label{dynmatr5c}
\end{equation}
The latter form may be thus considered as a Schmidt decomposition
 (\ref{VSchmidt}) of the superoperator $L$ for a CP--map, since the
non-negative eigenvalues of $D$ are simultaneously
Schmidt coefficients of $L=D^R$.

Alternatively one may find a matrix $\bf A$ of size $N^2 \times k$
such that $D_{\Phi}={\bf AA}^{\dagger}$.
This is always possible since $D$ is positive.
 Then the operator $A^{(i)}$ is
obtained by reshaping the $i$--th column of $\bf A$ into a square matrix.
Using the eigen representation of $D_{\Phi}$
the quantum map $\Phi$ defined in (\ref{dynmatr1}) becomes
\begin{equation}
\rho'_{m\mu}=
D_{\stackrel{\scriptstyle m n}{\mu \nu}} \rho_{n \nu}=
\sum_{i=1}^{k} A^{(i)}_{m n} \rho_{n\nu}
{\bar A}^{(i)}_{\mu \nu}
 \label{dynmatr6b}
\end{equation}
and it may be written in the {\sl canonical Kraus form}
\begin{equation}
\rho' =
\sum_{i=1}^{k} d_i \ \chi^{(i)} \rho (\chi^{(i)})^{\dagger}=
\sum_{i=1}^{k} A_i \rho A_i^{\dagger}.
 \label{dynmatr6}
\end{equation}
Let us compute the matrix $E \equiv \sum_{i=1}^{k} A_i^{\dagger}A_i$,
\begin{equation}
E_{\mu\nu}=
\sum_{i=1}^{k} {\bar A}^{(i)}_{m\mu}  A^{(i)}_{m\nu} =
\sum_{i=1}^{k} d_i {\bar \chi}^{(i)}_{m\mu}
 \chi^{(i)}_{m\nu}=
 D_{\stackrel{\scriptstyle m\mu}{m\nu}},
 \label{dynmatr7}
\end{equation}
and therefore  $ E={\rm Tr}_A D_{\Phi}$.
Due to the trace preservation constraint (\ref{dynmatr3c})
$E={\mathbbm 1}$, so
the operators $A_i$ satisfy the completeness relation
(\ref{Kraus2}). Hence (\ref{dynmatr6})
is equivalent to the Kraus form (\ref{Kraus1})
and represents a trace preserving CP map. We have thus shown that
any positive dynamical matrix $D$,
which satisfies the condition (\ref{dynmatr3c}),
specifies uniquely a quantum operation,
Hence any $N$-positive map
$\Phi$ leads to a positive matrix $D_{\Phi}$, so
$\Phi$ is completely positive.

On the other hand,
the Kraus representation (\ref{Kraus1})
is not unique -- two sets of the Kraus operators
$A_i, i=1,...,l$ and $B_j,j=1,...,n$
represent the same operation
if and only if the dynamical matrices given by
(\ref{dynmatr5b}) are equal.
This is the case if there exists a unitary matrix
$\bf V$ of size $mN$ such that
${\bf A}={\bf BV}$
so that the dynamical matrices both sets generate,  are equal,
\begin{equation}
D={\bf AA}^{\dagger}={\bf BV(BV)}^{\dagger}={\bf BB}^{\dagger}
 \label{dynmatr9}
\end{equation}
Here $m={\rm max}\{l, n\}$
and the shorter list of the Kraus operators is formally extended
by $|l-n|$ zero operators.

In principle the number of Kraus operators
in (\ref{Kraus1}) may be arbitrarily large.
 However, for any operation
one may find its Kraus form
 consisting of not more operators
than the rank $k$ of the dynamical matrix $D$. 
This {\sl Kraus rank} will never exceed the dimension of the dynamical
matrix equal to $N^2$.
Moreover, the Kraus operators $A_i$ may be chosen to be orthogonal
\cite{Ha03,AA03}.
To find such a {\sl canonical Kraus form}
given by (\ref{dynmatr6})
for an arbitrary operation $\Phi$ it is enough to find
its linear matrix $L$, reshuffle it to obtain the dynamical
matrix $D_{\Phi}$, diagonalize it, and out of its
eigenvalues $d_i$ and reshaped eigenvectors $|\chi_i\rangle$
construct by (\ref{dynmatr5})
the orthogonal Kraus operators $A_i$. They satisfy
\begin{equation}
\langle A_i | A_j \rangle = {\rm Tr} A_i^{\dagger} A_j
= \sqrt{d_i d_j} \langle \chi_i |\chi_j \rangle
=d_i \delta_{ij}.
 \label{dynmatr8}
\end{equation}

In a sense the Kraus form (\ref{Kraus1}) of a quantum operation $\Phi$
can be compared with an arbitrary decomposition of a density matrix,
$\rho=\sum_i p_i |\phi_i\rangle \langle \phi_i|$, while
its eigen decomposition
corresponds to the
canonical Kraus form of the map,
for which $||A_j||^2=d_j$.
In the generic case
of a nondegenerate dynamical matrix $D$, the canonical Kraus form
is specified uniquely, up to free phases, which may
be put in front of each Kraus operator $A_i$.

As discussed in section \ref{sec:jamiol}
such an analogy between the quantum maps (dynamics) and the quantum states
(kinematics) may be pursued much further. Let us state at this point
that the dynamical matrix is a linear function of the quantum operations
$\Phi$ and $\Psi$ in the sense that 
\begin{equation}
 D_{a \Phi + b \Psi} =a D_{\Phi} + b D_{\Psi} .
 \label{dynmatr10}
\end{equation}

An arbitrary quantum operation $\Phi$
is uniquely  characterized by any of the two matrices
$L$ or $D_{\Phi}=L^R$, but the meaning of their spectra is
entirely different. The dynamical
matrix $D_{\Phi}$ is Hermitian, while $L$ is
not and in general its eigenvalues $z_i$ are complex.
Let us order them according to their moduli,
$|z_1|\ge |z_2|\ge \cdots |z_{N^2}|\ge 0$.
A continuous linear operation $\Phi$ sends the convex compact set ${\cal
M}^{(N)}$ into itself. Therefore, due to the fixed--point theorem,
this transformation has a fixed point  --
an invariant state  $\sigma_1$ such that $L\sigma_1=\sigma_1$.
Thus $z_1=1$ and all eigenvalues fulfill $|z_i|\le 1$,
since otherwise the assumption that $\Phi$ is positive would be violated
\cite{TV00}.

\begin{table}
\caption{Quantum operations $\Phi:{\cal M}^{(N)}\to {\cal M}^{(N)}$:
properties of superoperator $L$  \mbox{and dynamical matrix $D_{\Phi}=L^R$}}
  \smallskip
\hskip -0.3cm
{\renewcommand{\arraystretch}{1.67}
\begin{tabular}
 [c]{||c|c|c||}
\hline \hline
Matrices & Superoperator $L=D^R$ & Dynamical matrix $D_{\Phi}$ \\
 \hline
Hermicity & No & Yes \\
\hline
Trace   &
\parbox{4.2cm}{\centering spectrum is symmetric \\
                       $ \Rightarrow$ \quad
                    tr$L \in {\mathbb R}$ }
& tr$D_{\Phi}=N$ \\
\hline
\parbox{2.5cm}{\centering (right) \\ Eigenvectors} &
\parbox{4.2cm}{\centering invariant states \\
                    or transient corrections } &
Kraus operators \\
\hline
Eigenvalues &
\parbox{4.2cm}{\centering $|z_i| \le 1$ \\
              $-\ln |z_i|$ -- decay rates} &
\parbox{4.2cm}{\centering weights of Kraus \\
                   operators, $d_i \ge 0$   }  \\
\hline
\parbox{2.8cm}{\centering Unitary evolution \\ $D_{\Phi}=(U\otimes U^*)^R$}
 & $||L||_2=N$ & $ S({\vec d}')=0$ \\
\hline
Coarse graining & $||L||_2=\sqrt{N}$ & $ S({\vec d}')= \ln N$ \\
\hline
\parbox{2.8cm}{\centering Complete depola-\\ risation, $D_{\Phi}={\mathbbm 1}$}
 & $||L||_2=1$ & $S({\vec d}')=2\ln N$ \\
\hline \hline
\end{tabular}
}
\label{tab:dynmat}
\end{table}
The trace preserving condition applied to the
eigen equation  {\mbox{ $L\sigma_i= z_i\sigma_i$}},
implies that if $z_i \ne 1$ than
tr$(\sigma_i) = 0$.
If $r=|z_2|<1$ then the matrix $L$ is
{\sl primitive} \cite{MO79} and all states converge to the
invariant state $\sigma_1$.
 If $L$ is diagonalizable
(there is no degeneracy in the spectrum or
  there exists no nontrivial blocks in
the Jordan decomposition of $L$, so that the number of right eigenvectors
$\sigma_i$ is equal to the size of the matrix $N^2$),
then any initial state $\rho_0$  may be
expanded in the eigenbasis of $L$,
\begin{equation}
\rho_0=\sum_{i=1}^{N^2}
     c_i \sigma_i
{\rm \quad \quad while \quad \quad }
 \rho_t=L^t\rho_0=\sum_{i=1}^{N^2}
     c_i z_i^t \sigma_i.
 \label{matrixL1}
\end{equation}
Therefore $\rho_0$ converges exponentially fast to the invariant state
$\sigma_1$ with the decay rate not smaller than $-\ln r$
and the right eigenstates $\sigma_i$ for $i\ge 2$ play the role of the
transient traceless corrections to $\rho_0$. The super-operator $L$
sends Hermitian  density matrices into Hermitian density matrices,
$\rho_1^{\dagger}=\rho_1=L\rho_0=L\rho_0^{\dagger}$, so
\begin{equation}
{\rm if} \quad \qquad
L\chi= z \chi
{\rm \quad \quad then \quad  \quad}
L\chi^{\dagger}=z^* \chi^{\dagger},
 \label{matrixL2}
\end{equation}
and the spectrum of $L$ (contained in the unit circle)
is symmetric with respect to the real axis.
Thus the trace of $L$ is real, as follows also from
the Hermicity of $D_{\Phi}=L^R$.

On the other hand, the real eigenvalues $d_j$ of the dynamical
matrix
$D_{\Phi}$ satisfy the normalization condition (\ref{dynmatr3c}),
 $\sum_j d_j=N$, which we assume to be finite. If the map $\Phi$
is completely positive all eigenvalues $d_j$ of $D_{\Phi}$ are non--negative,
and the matrix $\rho_{\Phi} \equiv D_{\Phi}/N$ may be interpreted as a density
matrix acting in ${\cal H}_{N^2}$.
The eigenvalues of $\rho_{\Phi}$, equal to
${\vec d}'={\vec d}/N$, determine
the weights of different operators $A_j$
contributing to the canonical Kraus form of the map
given by (\ref{dynmatr5}).
To characterize this probability vector quantitatively we
use the Shannon  entropy  to define
the {\sl entropy of an operation} $\Phi$,
\begin{equation}
S(\Phi)\equiv  S(\frac {1}{N}{\vec d}) =
-\sum_{i=1}^{N^2} (d_i/N) \ln(d_i/N) =  S_N(\rho_{\Phi}),
\label{entoper}
\end{equation}
equal to the von Neumann entropy of $\rho_{\Phi}$.
In order  to  characterize,
 to what extent the distribution of the elements
of the vector $\vec d$ is uniform one may also
use other quantities like linear entropy,
participation ratio or generalized R{\'e}nyi entropies.
If $S({\vec d}')=0$, the dynamical matrix $D_{\Phi}$ is of rank one,
so the state $\rho_{\Phi}$ is pure. Then Eq. (\ref{dynmatr5b}) reduces
to $D_{\stackrel{\scriptstyle m\mu}{n\nu}} =  U_{m\mu}U^*_{\nu n}$,
while $L=U\otimes U^*$
and the map $\Phi$ represents a unitary rotation.
The larger the entropy $S$ of an operation, the more
terms enter effectively into the canonical Kraus form, and the
larger are effects of decoherence in the system.

For any finite $N$ the entropy is bounded by $S_{max}=2\ln N$,
which is achieved for the rescaled identity matrix,
 $D={\mathbbm 1}_{N^2}/N$. This matrix represents the
{\sl completely depolarizing channel}
$\Phi_*$, such that any initial state $\rho$
is transformed into the maximally mixed state,
\begin{equation}
\Phi_*(\rho)=\rho_*={\mathbbm 1}_N/N
{\rm \quad for \quad any \quad state \quad} \rho \ . 
\label{depolchan}
\end{equation}
Under the action of this map complete decoherence takes place
already after the first iteration.
It is easy to see that the spectrum of the corresponding superoperator
$L_*=\frac{1}{N}({\mathbbm 1}_N)
^R$ consists of
$z_1=1$ and $z_i=0$ for $i \ge 2$.
In general the norm of the superoperator,
$||L||_2={\sqrt{{\rm Tr} LL^{\dagger}}}$,
 may thus be considered as
another quantity characterizing the decoherence induced by the map.
The norm $||L||_2$
varies from unity for the completely
depolarizing channel $\Phi_*$ (total decoherence) to $N$ for
any unitary operation (no decoherence).
Both quantities are connected,
since $||L||_2^2$,
equal to $||D||_2^2={\rm tr}D^2$,
is a function of the R{\'e}nyi entropy of order $\alpha=2$
of the spectrum of $D$ rescaled by $1/N$.
Making use of the monotonicity of the R{\'e}nyi entropies
\cite{BS93} we get
$||L||_2 \ge N \exp \bigr( -S(\Phi)/2\bigl)$.

An alternative way to characterize the properties
of a superoperator is to use its trace norm,
\begin{equation}
\gamma \equiv  ||L||_1=
{\rm Tr} | L|=
{\rm Tr} \sqrt{LL^{\dagger}} =
\sum_{i=1}^{N^2}  \xi_i ,
 \label{normL1}
\end{equation}
where $\xi_i$ are the singular values of $L=D^R$.
As discussed  in \cite{HHH02,CW03}
this quantity is useful to determine separability of the
state associated with the dynamical matrix, $\rho_{\Phi}=D/N$.

If the set of Kraus operators
$\{A_i\}_{i=1}^k$ determines a quantum operation $\Phi$
then for any two unitary matrices $V$ and $W$ of size $N$
the set of operators $A_i'=VA_iW$ satisfies
the relation (\ref{Kraus2}) and defines the operation
\begin{equation}
\rho\to \rho'={\Phi}_{VW}(\rho)=
\sum_{i=1}^k A'_i\rho A_i'^{\dagger} =
  V\Bigl( \sum_{i=1}^k A_i(W\rho W^{\dagger})
   A_i^{\dagger}\Bigr) V^{\dagger}.
\label{Krausprim}
\end{equation}
The operations $\Phi$ and $\Phi_{VW}$ are in general different, but {\sl
unitarily similar}, in a sense that the spectra of the dynamical
matrices are equal.
Thus their generalized entropies are equal, so
$S(\Phi)=S(\Phi_{VW})$ and $||L||_2=||L_{VW}||_2$. The latter
equality follows directly from the following law
of the transformation of superoperators,
\begin{equation}
 L_{VW}= (V\otimes V^*) L (W\otimes W^*),
\label{LLprim}
\end{equation}
which is a consequence of (\ref{dynmatr5c}).

\section{Unital \& bistochastic maps}
\label{sec:bistoch}

Consider a completely positive quantum
operation $\Phi$  defined by Kraus operators (\ref{Kraus1}).
If the set of $k$ operators $A_i$
satisfies the relation,
\begin{equation}
 \sum_{i=1}^k A_i A_i^{\dagger} = {\mathbbm 1}
=  {\rm tr}_B D
 \label{Kraus3}
\end{equation}
the operation is {\sl unital} (or {\sl exhaustive}),
 i.e. the maximally mixed state $\rho_*={\mathbbm 1}/N$
remains invariant, $\Phi(\rho_*)=
\frac{1}{N}\sum_{i=1}^k A_i A_i^{\dagger} = \rho_*$.
Since $(\rho_*)_{mn}=\frac{1}{N}\delta_{mn}$
then  the elements of $\Phi(\rho_*)$ are
$\frac{1}{N}
 L_{\stackrel{\scriptstyle m\mu}{n\nu}} \delta_{n\nu}=
\frac{1}{N}
 D_{\stackrel{\scriptstyle m n}{\mu n}}$,
which explains the right hand side equality in (\ref{Kraus3}),
related to the properties of the dynamical matrix $D$.

Observe the similarity between the condition (\ref{Kraus2}) for the
preservation  of trace and the unitality constraint (\ref{Kraus3}).
Unitality imposes that the sum of all elements
in each row of the corresponding matrix
$M$ defined by (\ref{KrausMM}) is equal to one.
Hence for any map $\Phi$ which is simultaneously trace preserving and
unital, the Kraus matrix $M$ is doubly--stochastic (bistochastic).
Therefore a trace preserving, unital completely positive map is
called a {\sl bistochastic map}
\cite{LS93,AHW00},
and may be considered as a noncommutative analogue
of the action of a bistochastic matrix $B$ on a probability vector -- see
table \ref{tab:bisto}. In the former case,
the maximally mixed state $\rho_*$ is $\Phi$--invariant,
while the uniform probability vector is an invariant
vector of $B$.

If all Kraus operators are Hermitian, $A_i=A_i^{\dagger}$,
the channel is bistochastic
but this is not a necessary condition.
For example,
unitary evolution may be considered as the
simplest case of the bistochastic map with $k=1$.
A more general class of bistochastic channels is given by
a convex combination of unitary operations, also
called {\sl random  external fields} (REF) \cite{AL87},
\begin{equation}
\rho'= \Phi(\rho)=\sum_{i=1}^k p_i V_i\rho V_i^{\dagger},
\quad {\rm with} \quad
p_i >0 \quad {\rm and} \quad
\sum_{i=1}^k p_i =1,
 \label{exterrandfield}
\end{equation}
where each operator $V_i$ is unitary.
The Kraus form (\ref{Kraus1})
can be reproduced setting $A_{i}=\sqrt{p_i} V_i$.

In general, a map for which all
Kraus operators are normal, $[A_i,A_i^{\dagger}]=0$,
is bistochastic.
Any convex combination of two bistochastic
maps is bistochastic, and similarly,
any convex combination of two random external fields
belongs to this class. Thus the spaces
of bistochastic maps and the spaces of random external
fields are convex.
For $N=2$ both sets coincide and any bistochastic map
may be represented as a combination
of unitary operations - see Fig. \ref{fig:oper3b}a.
Such one--qubit bistochastic maps are also called {\sl Pauli channels}.
 For $N\ge 3$ the set of bistochastic maps
  ${\cal B}_N$ is larger than the set of  REFs \cite{LS93}.

For any quantum channel $\Phi(\rho)$
one defines its {\sl dual channel} ${\tilde{\Phi}}(\rho)$,
such that the Hilbert--Schmidt scalar product satisfies
$\langle \Phi(\sigma) |\rho\rangle =
\langle \sigma |{\tilde{\Phi}}(\rho)\rangle$
 for any states $\sigma$ and $\rho$.
If a CP map is given by the Kraus form
 $\Phi(\rho)=\sum_i  A_i \rho A_i^{\dagger}$,
the dual channel reads
${\tilde {\Phi}}(\rho)=\sum_i  A_i^{\dagger}\rho A_i$.
Making use of (\ref{dynmatr5b}) we obtain a link between
the dynamical matrices representing dual channels,
\begin{equation}
L_{\tilde{\Phi}}= (L_{\Phi}^T)^F= (L_{\Phi}^F)^T
\quad \quad {\rm and} \quad \quad
D_{\tilde{\Phi}}= (D_{\Phi}^T)^F=(D_{\Phi}^F)^T={\overline{D_{\Phi}^F}} .
 \label{dualDL}
\end{equation}
Since neither the transposition nor
the flip  modify the spectrum of a matrix,
the spectra of the dynamical matrices
for dual channels are the same, as well as their entropies,
 $S(\Phi)=S({\tilde \Phi})$.

Comparing the conditions  (\ref{Kraus2}) and (\ref{Kraus3})
we see that if channel $\Phi$ is trace preserving, its dual
$\tilde \Phi$ is
unital, and conversely,
if channel $\Phi$ is unital then $\tilde \Phi$ is trace preserving.
Thus the channel dual to a bistochastic one is bistochastic.

For any quantum operation $\Phi$ given by the Kraus form (\ref{Kraus1})
one defines its  {\sl effect}  by $E \equiv \sum_{i=1}^k A_i A_i^{\dagger}$.
Due to Eq. (\ref{Kraus3})
the effect is obtained by partial trace of the dynamical matrix,
$E={\rm Tr}_B D$, so it is Hermitian and positive.
Since $E(\Phi)=\frac{1}{N} {\tilde {\Phi}}(\rho_*)$,
it is clear that the effect of any operation does not depend of
the particular choice of the Kraus operators used to represent it.
For bistochastic maps  $\Phi(\rho_*)=\rho_*$ and
 $E={\mathbbm 1}$.

Bistochastic channels are the only ones which do not decrease the von
Neumann entropy of any state they act on. To see
this consider the image of the maximally mixed state,
with maximal entropy $\ln N$. If the map is not unital,
(the channel is not bistochastic),
then $\rho'=\Phi(\rho_*)$ differs from
$\rho_*$, so its entropy decreases.
As an important example of bistochastic channels consider
the {\sl coarse graining} operation, which sets all
off-diagonal elements of a density matrix to zero,
 $\Psi_{\rm CG} (\rho)={\rm diag}(\rho)$.
It is described by the diagonal dynamical matrix,
$D^{\rm CG}_{\stackrel{\scriptstyle m \mu}{n
\nu}}=\delta_{mn}\delta_{\mu\nu}\delta_{m\mu}$,
with $N$ elements equal to unity.

\bigskip
\begin{table}
\caption{Quantum maps acting on density matrices and
  given by positive definite dynamical matrix $D\ge 0$ versus
  classical Markov dynamics on probability vectors defined by
transition matrix $T$ with non-negative elements}
 \smallskip
\hskip -0.5cm
{\renewcommand{\arraystretch}{1.43}
\begin{tabular}
[c]{||c c|c c||}
\hline \hline
Quantum  &
\parbox{4.4cm}{\centering completely \\ positive maps: } &
Classical &
\parbox{4.4cm}{\centering Markov chains \\ given by: } \\
\hline
 $S_1^Q$ & Trace preserving, Tr$_A D={\mathbbm 1}$ &
 $S_1^{Cl}$ &
 Stochastic matrices $T$  \\
 \hline
$S_2^Q$ & Unital, Tr$_B D={\mathbbm 1}$   &
 $S_2^{Cl}$ & $T^T$ is stochastic \\
 \hline
$S_3^Q$ & Unital \& trace preserving maps &
 $S_3^{Cl}$ & Bistochastic matrices $B$ \\
 \hline
$S_4^Q$ &
\parbox{4.4cm}{\centering Maps with $A_i=A_i^{\dagger}$ \\
                 $ \Rightarrow D=D^T$ } &
  $S_4^{Cl}$ &
  \parbox{4.4cm} {\centering Symmetric bistochastic \\
                           matrices, $B=B^T$} \\
 \hline
$S_5^Q$ & \parbox{4.4cm}{\centering Unistochastic
operations,
\\  $D=U^R (U^R)^{\dagger}$ }  & $S_5^{Cl}$ &
\parbox{4.4cm}{\centering  Unistochastic matrices, \\
 $B_{ij}=|U_{ij}|^2$ }  \\
 \hline
$S_6^Q$ &
\parbox{4.6cm}{\centering Orthostochastic operations,
\\ $D=O^R (O^R)^T $ } & $S_6^{Cl}$ &
\parbox{4.4cm}{\centering  Orthostochastic matrices,
\\ $B_{ij}=|O_{ij}|^2$ }  \\
 \hline
$S_7^Q$ &
\parbox{4.8cm}{\centering Unitary transformations} & $S_7^{Cl}$ &
\parbox{4.4cm}{\centering  Permutations} \\
\hline \hline
\end{tabular}
}
\label{tab:bisto}
\end{table}

Let us analyze in some detail the set  ${\cal BU}_N$ of
{\sl unistochastic} {\sl operations},
for which there exists the representation (\ref{unitevol5}).
The initial state of the environment is maximally mixed,
$\sigma=\rho_*={\mathbbm 1}/N$, so the quantum  map $\Psi_U$ is
determined by a unitary  matrix $U$ of size $N^2$.
The reshaped Kraus operators $A_i$ are
proportional to the eigenvectors (\ref{dynmatr5})
of the dynamical matrix $D_{\Psi_U}$. On the other hand, they enter
also the Schmidt decomposition (\ref{VSchmidt})
of $U$ as shown in (\ref{unitevol5}),
 and are proportional to the
eigenvectors of  $(U^R)^{\dagger} U^R$.
Therefore
\begin{equation}
 D_{\Psi_U}= \frac{1}{N}  (U^R)^{\dagger} U^R
{\quad \rm so \quad that \quad}
L= \frac{1}{N} \bigl[(U^R)^{\dagger}U^R \bigr]^R .
\label{unisto1}
\end{equation}

We have thus arrived at an important result:
for any unistochastic map
the spectrum of the
dynamical matrix $D$ is given  by the
Schmidt coefficients, $d_i=\lambda_i/N$,
of the unitary matrix $U$ treated as an element of the composite
HS space. Hence the entropy of the operation $S(\Psi_U)$ is equal
to the entanglement entropy of the unitary matrix, $S(U)$.

Moreover, the linear entropy of entanglement of $U$
studied in \cite{Za01} is a function of the norm
of the superoperator, $E(U)=1-\sum_i \lambda_i^2=1-||L||_2^2/N^2$.
It vanishes for any local operation, $U=U_1 \otimes U_2$,
for which the superoperator is unitary,
$L=U_1 \otimes  U_1^*$ so $||L||_2^2=N^2$.
The resulting unitary operation is an isometry,
and can be compared with a permutation $S^{\rm Cl}_7$ acting on the
simplex $\Delta_{N-1}$ of classical probability vectors.
If the matrix $U$ is
orthogonal the corresponding dynamical matrix
is symmetric, $D_{\Psi_O}=(O^R)^T O^R=D_{\Psi_O}^T$.
The corresponding operation will be called an {\sl orthostochastic
map},
as listed in Table \ref{tab:bisto}.
The spaces listed there satisfy the following relations
$S_1 \cap S_2=S_3$ and
$S_3 \supset S_5 \supset S_6$ and
$S_5 \supset S_7$ in both, classical and quantum set ups.
However, the analogy is not exact: the inclusions
$S^{\rm Cl}_4 \subset S^{\rm Cl}_3$ and
 $S^{\rm Cl}_7 \subset S^{\rm Cl}_6$ and
 $S^{\rm Q}_6 \subset S^{\rm Q}_4$ do not have their counterparts.
Note that unitary (orthogonal) matrices
defining quantum maps
 ${\cal M}^{(N)} \to {\cal M}^{(N)}$
 are of size $N^2$ while these determining Markov chains
 $\Delta_{N-1} \to \Delta_{N-1}$ are of size $N$.
Note that
already for $N=2$
not all bistochastic maps are unistochastic \cite{MKZ04}.

\smallskip

\section{One qubit maps}
\label{sec:onequibit}

In the simplest case $N=2$ the
quantum operations are called {\sl binary channels}.
In general, the space ${\cal CP}_N$ is $N^4-N^2$ dimensional.
Hence the space ${\cal CP}_2$ of binary channels
has $12$ dimensions and there exist several ways to
parametrise it  \cite{FA99,KR01}.
The dynamical matrix provides a straightforward,
but not very transparent method to achieve this goal.
Any  Hermitian matrix $D$ of size $4$,
which satisfies tr$_A D={\mathbbm 1}$
 may be written as
\begin{equation}
D =\left[
\begin{array}{cc|cc}
\frac{1}{2}+a & x               & y &  z \\
{\bar x}        & \frac{1}{2}-b  & w & -u \\
\hline
{\bar y} & {\bar w} & \frac{1}{2}-a  & -x \\
{\bar z} & -{\bar u} & -{\bar x} & \frac{1}{2}+ b
\end{array} \right] ,
\label{densaDD}
\end{equation}
where $a$ and $b$ are real and  $x,y,z,u,w$
complex parameters. If they are adjusted in such a way to
assure positivity of $D$,
then this matrix represents a trace preserving CP map.
Note that $D/N$ represents a certain density matrix in ${\cal M}^{(4)}$.
If additionally $a=b$ and $y=u$ then
the condition tr$_B D={\mathbbm 1}$ is fulfilled,
so the map is bistochastic.

\vskip 0.1cm
\begin{figure} [htbp]
   \begin{center}
\
 \includegraphics[width=12.3cm,angle=0]{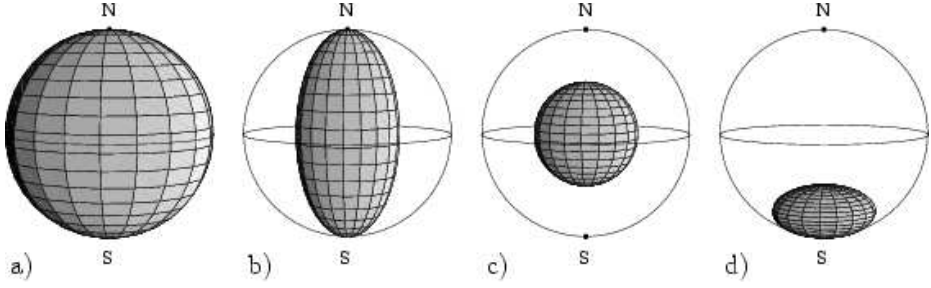}
\caption{One qubit operations acting on the Bloch ball (a),
 bit flip with $p=0.5$ (b), depolarizing channel with $p=0.5$ (c)
and  a non--unital operation of amplitude damping
with $p=0.75$ (d).}
 \label{fig:oper3}
\end{center}
 \end{figure}

An alternative approach to the problem
is obtained using the Stokes para\-me\-tri\-sation,
which involves Pauli matrices $\sigma_k$.
Any state
$\rho= ({\mathbbm 1}+ {\vec \tau}\cdot {\vec \sigma})/2$,
is characterized by the Bloch vector $\vec \tau$.
If $\rho'=\Phi(\rho)$ is represented by $\vec \tau'$, then
any linear dynamics inside the Bloch ball
may be described by an affine transformation of the Bloch vector
\begin{equation}
{\vec \tau}'=t{\vec \tau} + {\vec \kappa},
\label{matrTtau}
\end{equation}
where $t$ denotes a real matrix of size $3$,
while $\vec \kappa=(\kappa_x,\kappa_y,\kappa_z)$
is the translation
vector. It determines the image of the ball center,
$\Phi(\rho_*)=({\mathbbm 1} + {\vec \kappa} \cdot {\vec
\sigma})/2$.
For unital maps $\Phi(\rho_*)=\rho_*$ so ${\vec \kappa}=0$.
Any positive map sends the Bloch ball into an ellipsoid
which may degenerate to an ellipse, an interval or a point.
Out of $12$ real parameters in (\ref{matrTtau})
$3$ eigenvalues of $tt^{\dagger}$ determine the size of the ellipsoid, $6$
related to the eigenvectors specify the orientation of its axis
and the $3$ parameters of the vector $\vec \kappa$
characterize the position of its center.

The real matrix $t$ may be brought to the form, $t=O_1 \eta O_2$,
where $O_i\in O(3)$ and the diagonal matrix $\eta$ contains
non-negative singular values of $t$.
However, it is convenient to impose an extra condition that
both orthogonal matrices represent proper
rotations ($O_i \in SO(3)$),
at the expense of allowing some components
of the diagonal vector $\vec \eta$ to be negative.  This decomposition
is not unique: the moduli $|\eta_i|$ are equal to singular values of
$t$ and the
sign of the product $\eta_x\eta_y\eta_z={\rm det}(t)$ is fixed.
However, the signs of any two components of $\vec \eta$
may be changed by conjugating the map with a Pauli matrix.

In general, for any one--qubit CP map $\Phi$
one may find unitary matrices $U$ and $W$ of size $2$
such that
the unitarily similar operation ${\Phi}_{UW}$ defined by
(\ref{Krausprim}) is represented by the diagonal matrix $\eta$.
For simplicity we may thus restrict ourselves to
the maps for which the matrix $t$ is diagonal
and consists of the damping (distortion) vector,
 ${\vec{\eta}}=(\eta_x,\eta_y,\eta_z)$.
In this case the ellipsoid has the form
\begin{equation}
 \Bigl( \frac{ x-\kappa_x}{\eta_x}\Bigl)^2 +
 \Bigl( \frac{ y-\kappa_y}{\eta_y}\Bigl)^2 +
\Bigl( \frac { z-\kappa_z}{\eta_z}\Bigl)^2 = 1.
\label{elipstau}
\end{equation}
The affine transformation (\ref{matrTtau})
determines the superoperator $L$ of the map.
Reshuffling the superoperator matrix according to (\ref{dynmatr3})
we obtain the dynamical matrix $D=L^R$ which corresponds to the map
$\Phi_{{\vec \eta},{\vec \kappa}}$,
\begin{equation}
D = \frac{1}{2}
\left[
\begin{array}{cccc}
  1+\eta_z+\kappa_z  & 0 & \kappa_x+i\kappa_y  & \eta_x+\eta_y  \\
  0      & 1-\eta_z+\kappa_z & \eta_x-\eta_y&\kappa_x+i\kappa_y  \\
\kappa_x-i\kappa_y & \eta_x -\eta_y & 1-\eta_z-\kappa_z & 0   \\
\eta_x+\eta_y & \kappa_x - i \kappa_y &   0 &1+\eta_z-\kappa_z
\end{array} \right] ,
\label{densaDtau}
\end{equation}
 clearly a special case of (\ref{densaDD}).
For unital maps ${\vec \kappa}=0$ and the matrix $D$ splits into two
blocks and its eigenvalues are
\begin{equation}
d_{0,3}=\frac{1}{2}[1+\eta_z\pm(\eta_x+\eta_y)]
{\quad \rm and \quad}
d_{1,2}=\frac{1}{2}[1-\eta_z\pm(\eta_x-\eta_y)].
\label{etaddd}
\end{equation}
 Hence if the Fujiwara--Algoet conditions \cite{FA99}
\begin{equation}
(1\pm \eta_z)^2 \ge (\eta_x\pm \eta_y)^2
\label{postautau}
\end{equation}
are fulfilled, the dynamical matrix $D$ is positive definite
and the corresponding positive map $\Phi_{\vec \eta}$ is CP.
This condition shows that not all ellipsoids located inside the
Bloch ball may be obtained by acting on the ball with
a CP map - see recent papers
on one qubit maps \cite{FA99,TCDGS99,KR01,Oi01,Wo01,Uh01,RSW02,VV02}.

Note that dynamical matrices of unital maps do commute,
$[D_{\Phi_{\vec \eta}},D_{\Phi_{\vec \zeta}}]=0$.
Hence they share the same set of eigenvectors. 
When reshaped they form the orthogonal set of
Kraus operators consisting of the identity
$\sigma_0={\mathbbm 1}_2$, and the three Pauli matrices $\vec \sigma$.
Thus the canonical form of an arbitrary one--qubit bistochastic
map reads
\begin{equation}
\rho'=\Phi_{\vec \eta}(\rho) = \sum_{i=0}^3
d_i \sigma_i \rho \sigma_i,
\label{bistPauli}
\end{equation}
which explains the name {\sl Pauli channels}.

For concreteness let us distinguish some
one-qubit channels.
We are going to specify their
translation and distortion vectors,
$\vec \kappa$ and $\vec \eta$,
 since with use of the transformation (\ref{Krausprim})
we may bring the matrix $t$ to its diagonal form.
Alternatively, the channels may be defined using the
canonical Kraus form (\ref{dynmatr6})
in which Kraus operators are given by the eigenvalues and
eigenvectors of the dynamical matrix, $A_i=\sqrt{d_i}\chi_i$.
From (\ref{bistPauli}) it follows
that for any $N=2$ bistochastic map 
the Kraus operators are $A_i=\sqrt{d_i}\sigma_i$.
For the  {\sl decaying channel},
 (also called {\sl amplitude--damping channel}),
which is not bistochastic,
the eigenvectors of the dynamical matrix  give

\begin{equation}
 A_1=
 \left[ \begin{array}
 [c]{cc}
 1 & 0 \\
 0 & \sqrt{1-p}
 \end{array}  \right] 
{\rm \quad and \quad} 
  A_2=
 \left[ \begin{array}
 [c]{cc}
 0 & \sqrt{p} \\
 0 & 0
 \end{array}  \right]  \ . 
\label{vectdecay}
\end{equation}

Basic properties of some selected maps  are collected in
Table \ref{tab:2maps} and illustrated in Fig. \ref{fig:oper3}.

\bigskip
\begin{table}
\caption{Exemplary one--qubit channels:
distortion vector $\vec \eta$,
translation vector $\vec \kappa$ equal to zero for unital channels,
rescaled Kraus spectrum ${\vec d}'$, and the Kraus rank $k$.}
\smallskip
{\renewcommand{\arraystretch}{1.45}
\begin{tabular}
[c]{||c|c|c|c|c|c||}%
\hline \hline
{\bf Channels} & $\vec \eta$ & $\vec \kappa$ & unital &
${\vec d}^{\prime}$ & $k$\\
\hline
rotation & $(1,1,1)$ & $(0,0,0)$ & yes & $(1,0,0,0)$ & $1$\\
\hline
phase flip & $(1-p,1-p,1)$ & $(0,0,0)$ & yes & $(1-p/2,p/2,0,0)$ & $2$\\
\hline
decaying  & $(\sqrt{1-p},\sqrt{1-p},1-p)$ & $(0,0,p)$ & no & $(1-p/2,p/2,0,0)$
& $2$\\
\hline
depolarizing & $[1-x](1,1,1)$ & $(0,0,0)$ & yes & $\frac{1}{4}(4-3x,x,x,x)$ &
$4$\\
\hline
linear & $(0,0,q)$ & $(0,0,0)$ & yes &
\parbox{3.3cm}{\centering
 $\frac{1}{4}(1+q,1-q,$ \\ \ \ \ $1-q,1+q)$ }
& $4$ \\
\hline
planar & $(0,r,q)$ & $(0,0,0)$ & yes &
\parbox{3.3cm}{\centering
 $\frac{1}{4}(1+q+r,1-q-r,$ \\ \ \ \ $1-q+r,1+q-r)$ }
& $4$
 \\
\hline \hline
\end{tabular}
}
\label{tab:2maps}
\end{table}

\begin{figure} [htbp]
   \begin{center}
 \includegraphics[width=12.3cm,angle=0]{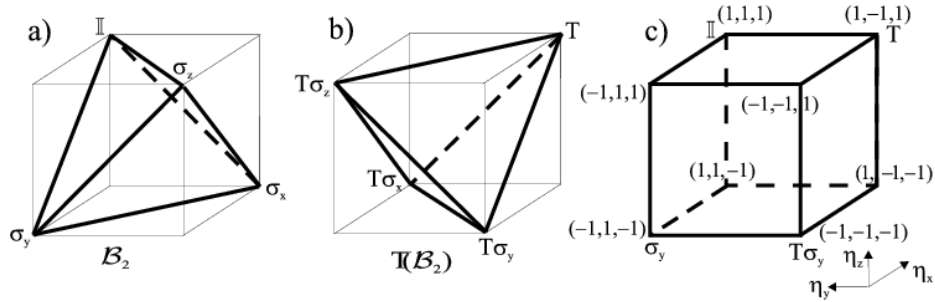}
 \vskip 0.1cm
\caption{Subsets of one--qubit maps:
a) set ${\cal B}_2$ of bistochastic maps
(unital \& CP),
  b) set $T({\cal B}_2)$ of unital \& CcP maps,
  c) set of positive (decomposable) unital maps.}
 \label{fig:oper3b}
\end{center}
 \end{figure}

The geometry of the set of bistochastic maps is simplest to understand
for $N=2$.
As discussed in section \ref{sec:bistoch} this set  coincides
 \cite{LS93} with the set of $N=2$
random external fields (\ref{exterrandfield}), which may be defined as
the convex hull of four unitary operations,
$\sigma_0={\mathbbm 1},\sigma_x,\sigma_y$ and $\sigma_z$.
Thus, in agreement with (\ref{bistPauli}),
any one-qubit bistochastic map $\Phi\in {\cal B}_2$
 may be written as a
convex combination of four matrices $\sigma_j$, $j=0,...,3$.
Since $\sigma_j=-i\exp(i\pi\sigma_j/2)$,
and the overall phase $-\pi$ is not relevant,
each Pauli matrix represents the rotation of the Bloch ball around
the corresponding axis by angle $\pi$.
The distortion vectors  ${\vec \eta}$ of four extremal
maps $\sigma_j$
read $(1,1,1)$,  $(1,-1,-1)$,   $(-1,1,-1)$, and
$(-1,-1,1)$, respectively. Hence,
in accordance with the constraints (\ref{postautau}),
the set of bistochastic maps ${\cal B}_2$ 
written in the canonical form 
(with diagonal matrix $t$),
forms a tetrahedron, see Fig. \ref{fig:oper3b}a.
Its center is occupied by the completely depolarizing channel
with ${\vec \eta}={\vec 0}$,
which may be expressed as a uniform mixture of four extremal
unitarities; $\Psi_*=({\mathbbm 1}+\sigma_x+\sigma_y+\sigma_z)/4$.
Interestingly, the set of non unital channels with a fixed
translation vector ${\vec \kappa}\ne 0$ forms a
convex set which resembles a tetrahedron with its corners rounded
\cite{RSW02}.

Observe that in the generic case $|\eta_i|<1$,
which suggests that a typical one-qubit CP map is a contraction.
This is true also for higher dimensions, $N>2$. 
The monotone distances
(e.g. the trace distance and the Bures distance \cite{NC00})
do not increase under the action of CP maps.
Apart from unitary (antiunitary) operations, for which
these distances are preserved,  the transformation
inverse to a quantum operation is not
an operation any more:
some mixed states are sent outside
the set of positive operators.

\section{Positive \& decomposable maps}
\label{sec:positiv}

Quantum transformations which describe physical processes are
represented
by completely positive maps. Why should we care about maps
which are not completely positive? On one hand it is
instructive to realize that seemingly innocent transformations
are not CP, and thus do not correspond to any physical process.
On the other hand maps which are positive, not completely positive
provide a crucial tool in the
investigation of entangled mixed states \cite{Per96,HHH96a,HHH01}.

Consider the transposition of a density matrix in a certain
 basis $T:\rho \to \rho^T$.
The corresponding superoperator $L_T$ entering (\ref{dynmatr1})
has the form $(L_T)_{\stackrel{\scriptstyle m \mu}{n \nu}}
=\delta_{m\nu}\delta_{n\mu}$,  and equals the dynamical matrix,
$L_T=L_T^R=D_T$. This is a permutation matrix which contains $N$
diagonal entries equal to unity and $N(N-1)/2$ blocks of size two,
Thus its spectrum, spec$(D_T)$, consists of $N(N+1)/2$
eigenvalues equal to unity and $N(N-1)/2$ eigenvalues equal to
$-1$, which is consistent with the constraint tr$D=N$.
The matrix $D_T$ is not positive, so the transposition $T$ is not
completely positive. Another way to reach this conclusion is to
act with the extended map of partial transposition on the maximally
entangled state (\ref{maxenta}) and to check that
$[T\otimes {\mathbbm 1}](| \psi\rangle\langle \psi|)$
has negative eigenvalues.

\begin{figure} [htbp]
   \begin{center}
\
 \includegraphics[width=7.5cm,angle=0]{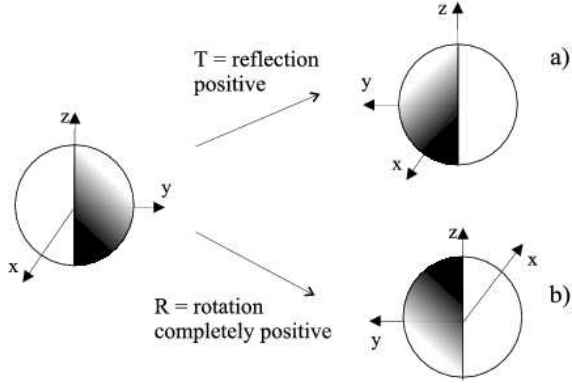}
\caption{Non-contracting transformations of the Bloch ball:
  a) transposition (reflection with respect to the $x-z$ plane) - not
completely positive; b) rotation by $\pi$ around $z$ axis - completely
positive.}
 \label{fig:oper4}
\end{center}
 \end{figure}

The transposition of an $N$--dimensional Hermitian matrix $D$,
 which changes
the signs of the imaginary part of the elements $D_{ij}$,
may be viewed as a reflection
 with respect to the $N(N+1)/2-1$
dimensional hyperplane, so the remaining $N(N-1)/2$
components do change their signs.
As shown in Fig.  \ref{fig:oper4}
this geometrical interpretation is simple to visualize
for $N=2$:
the transposition corresponds to a reflection of the Bloch ball with
respect to the $(x,z)$ plane - the $y$ coordinate changes its sign.
Note that rotation of the Bloch ball around
the  $z$-axis
by the angle $\pi$, realized by a unitary operation, $\rho\to
\sigma_z\rho\sigma_z$, also exchanges the 'western' and
the 'eastern' hemispheres, but is completely positive.

\vskip -0.3cm

As discussed in section \ref{sec:dynmat}
a given map $\Phi$ is not CP,
if the corresponding dynamical matrix $D$ contains
a negative eigenvalue. Let $m\ge 1$ denote
the number of the negative eigenvalues
(in short, the {\sl neg rank} of $D$).
Then the spectral decomposition of  $D$ takes the form
\begin{equation}
D=\sum_{i=1}^{N^2-m} d_i | \chi_i \rangle \langle \chi_i|  \  -
   \sum_{i=N^2-m+1}^{N^2} d_i| \chi_i \rangle \langle \chi_i| \ .
 \label{canonP}
\end{equation}
In analogy to the Kraus form (\ref{dynmatr6})
we may write the canonical form of a not completely positive map
\begin{equation}
\rho' =
\sum_{i=1}^{N^2-m} d_i \chi^{(i)} \rho (\chi^{(i)})^{\dagger}\ -
\sum_{i=N^2-m+1}^{N^2} d_i \chi^{(i)} \rho (\chi^{(i)})^{\dagger},
 \label{dynmatr6N}
\end{equation}
where the Kraus operators $A_i=\sqrt{d_i}\chi^{(i)}$
form an orthonormal basis.
The above form suggests that a positive map may be represented
as a {\sl difference} of two completely positive maps \cite{SS03}.
Even though this statement is correct (for finite $N$),
it does not solve the entire problem:
taking any two CP maps and constructing a quasi-mixture
(with negative weights allowed),
$\Phi=(1+a)\Psi_1^{\rm CP}-a\Psi_2^{\rm CP}$,
we do not know in advance how large the 
contribution $a$ of the negative part might be, to keep the map
$\Phi$ positive...
Although some criteria for positivity
are known for several years \cite{St63,Ja74,Ma75},
they do not allow one to perform a practical test,
whether a given map is positive.
A recently proposed technique of extending
the system  (and the map) a certain number of times is shown to
give a constructive test for positivity for a large class of maps
\cite{DPS03}.
In fact the characterization of the set ${\cal P}_N$ of positive maps:
${\cal M}^{(N)}\to {\cal M}^{(N)}$
is by far not simple and remains a subject of 
considerable mathematical interest
\cite{St63,Cho72,Wo76b,Cho80,TT83,Os92,Ky96,EK00,YH00,Ko00,MM01,LMM03,SS03,Ky03}.
By definition, ${\cal P}_N$
contains the set ${\cal CP}_N$ of all CP maps as its proper subset.

To learn more about the set of positive maps we will need
some other features of the operation of transposition
$T$. For any operation $\Phi$ the
modifications of the dynamical matrix induced by
a composition with $T$  may be described by the
partial transpose transformation 
\begin{equation}
L_{T{\Phi}}= L_{\Phi}^{F_1}, \quad D_{T{\Phi}}= D_{\Phi}^{T_A},
\quad {\rm and} \quad
L_{{\Phi}T}= L_{\Phi}^{F_2}, \quad D_{{\Phi}T}= D_{\Phi}^{T_B}.
 \label{trnasT}
\end{equation}
To demonstrate this it is enough to use the explicit form of
$L_T$ and the following law of composition of dynamical matrices.
Since the composition
 of two maps $\Psi \Phi$
results in the product of linear matrices,
\begin{equation}
L_{\Psi \Phi} =L_{\Psi} L_{\Phi},
{\quad \rm  hence \quad}
D_{\Psi \Phi} =[D_{\Psi}^R D_{\Phi}^R]^R .
 \label{composD}
\end{equation}
Even though the composition
 of two maps is usually written as $\Psi\cdot \Phi$,
to simplify the notation the symbol $\cdot$
will often be dropped. Note that both operations
commute, $\Psi \Phi =\Psi \Phi$, if
$[D_{\Phi}^R, D_{\Psi}^{R}]=0$.

Positivity of $D_{\Psi \Phi}$ follows also from
the fact that the composition of two CP maps is
completely positive. Alternatively one may
prove the following reshuffling lemma

{\sl Consider two Hermitian matrices $A$ and $B$ of the same size $KN$}.
\begin{equation}
{\it If \quad} A\ge 0
{\it \quad and \quad} B\ge 0
{\it  \quad then  \quad}
(A^RB^R)^R \ge 0.
\label{reshlem}
\end{equation}
It was formulated in a different 
 set up and proved in \cite{Ha03}.

\begin{figure} [htbp]
   \begin{center}
\
 \includegraphics[width=10.0cm,angle=0]{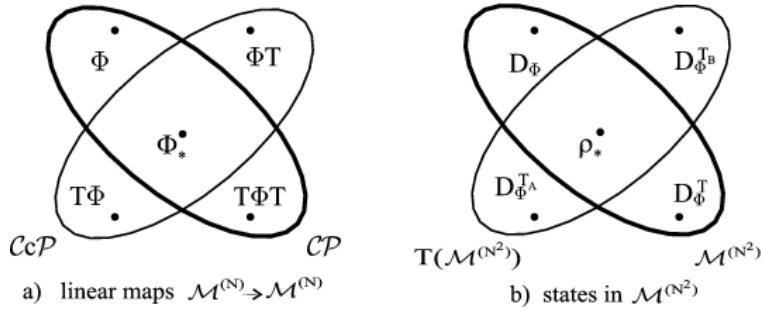}
\caption{a) Sketch of the set of CP maps
and of its image with respect to transposition: the set of CcP
maps, b) the isomorphic sets
of quantum states (dynamical matrices) and its image
under the action of partial transposition.}
 \label{fig:oper5}
\end{center}
 \end{figure}

The sandwiching of $\Phi$ between two
actions of transpositions
does not influence the spectrum of the dynamical matrix,
$L_{T{\Phi}T}= L_{\Phi}^{F}=L_{\Phi}^*$ and $D_{T{\Phi}T}=
D_{\Phi}^{T}=D_{\Phi}^*$. Thus if $\Phi$ is a CP map, so is
$T\Phi T$, (if $D_{\Phi}$ is positive so is $D_{\Phi}^T$)
 -- see Fig. \ref{fig:oper5}.

\smallskip

The non completely positive map $T$ of transposition allows
one to introduce the following  definition \cite{St63,Cho75a,Cho80}:

\smallskip
A map $\Phi$ is called {\sl completely co-positive} (CcP),
 if the map $T\Phi$ is CP.

\smallskip
Properties (\ref{trnasT}) of the dynamical matrix imply
that the map $\Phi T$ could be used instead to define the same set of
CcP maps. Thus any CcP map may be written in a Kraus--like form
\begin{equation}
\rho' = \sum_{i=1}^{k} A_i \rho^T A_i^{\dagger}.
 \label{CcoPkraus}
\end{equation}
Moreover, as shown in Fig. \ref{fig:oper5},
the set ${\cal C}{\rm c}{\cal P}$
 may be understood as the image
of $\cal CP$ with respect to the transposition.
Since we have already identified the transposition with a kind of
reflection, it is rather intuitive to observe that the set
${\cal C}{\rm c}{\cal P}$
is a twin copy of ${\cal CP}$ with the same shape and
volume. This property is easiest to analyze for the set ${\cal B}_2$ of
one qubit bistochastic maps \cite{Oi01}. The dual set of CcP  unital
one qubit maps, $T({\cal B}_2)$, forms a tetrahedron spanned by four
maps $T\sigma_i$ for $i=0,1,2,3$,
which {\sl is} the reflection of the set of the
bistochastic maps with
respect to the center of the tetrahedron - the completely depolarizing
channel $\Phi_*$ -see Fig. \ref{fig:oper3b}b. Observe that the
corners of ${\cal B}_2$ are formed by proper rotations,
for which det$(t)=\eta_1\eta_2\eta_3$ is equal to $+1$,
while the extremal points of the set of CcP maps represent
reflections for which det$(t)=-1$.

A positive map $\Phi$ is called {\sl decomposable}, if it may be expressed as
a convex combination of a CP map and a CcP map,
$\Phi=a \Phi_{\rm CP} + (1-a) \Phi_{\rm CcP}$
with $a\in [0,1]$. A relation between CP maps acting
on quaternion matrices and the decomposable maps
defined on complex matrices was shown by Kossakowski \cite{Ko00}.
An important characterization of the set ${\cal P}_2$ of positive maps
acting on (complex) states of one qubit
follows from the {\sl St{\o}rmer--Woronowicz theorem}
\cite{St63,Wo76b}

\smallskip
{\sl Every one-qubit positive  map $\Psi\in {\cal P}_2$
is decomposable.}
\smallskip

\begin{figure} [htbp]
   \begin{center}
\
 \includegraphics[width=12.3cm,angle=0]{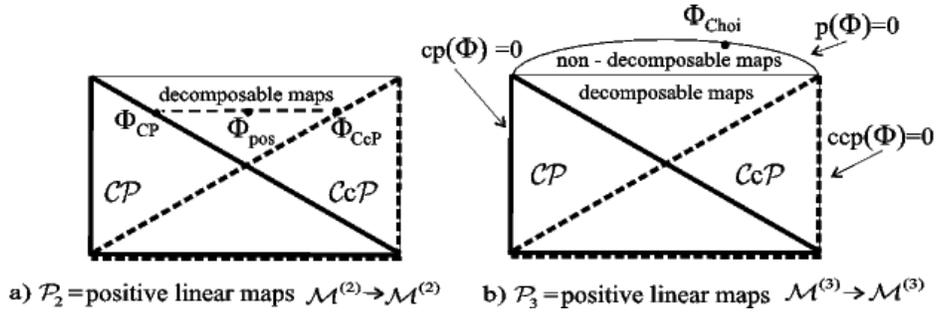}
\caption{Sketch of the set positive maps:
a) for $N=2$ all maps are decomposable,
b) for $N>2$ there exist non-decomposable maps.}
 \label{fig:oper6}
\end{center}
 \end{figure}

In other words, the entire set of one qubit positive maps can be
represented by the convex hull of the set of CP and CcP maps,
${\cal P}_2={\rm conv~~ hull~}\Bigl( {\cal CP}_2 \cup
{\cal C}c{\cal P}_2 \Bigr) $.
This property, illustrated in Fig. \ref{fig:oper6},
holds also for the maps ${\cal M}^{(2)}\to {\cal M}^{(3)}$
and
${\cal M}^{(3)}\to {\cal M}^{(2)}$ \cite{Wo76a},
but is not true in higher dimensions, in particular
in the sets  ${\cal P}_N$ with $N \ge 3$.
Consider a map defined on ${\cal M}^{(3)}$,
depending on three non-negative parameters,
\begin{equation}
\Psi_{a,b,c}(\rho)  \! = \!
\left[ \begin{array}
[c]{ccc}
a\rho_{11}+b\rho_{22}+c\rho_{33} & 0 & 0 \\
0 & c\rho_{11}+a\rho_{22}+b\rho_{33} & 0 \\
0 & 0 & b\rho_{11}+c\rho_{22}+a\rho_{33}
\end{array}  \right] -\rho .
\label{GenCHOI}
\end{equation}
The map  $\Psi_{2,0,2}\in {\cal P}_3$ was a first example of a
indecomposable map found by Choi in 1975 \cite{Cho75}
in connection with positive biquadratic forms.
As denoted  schematically in Fig.  \ref{fig:oper6}b
this map is extremal and belongs to the boundary of the convex set
${\cal P}_3$. The Choi map was generalized later in \cite{CL77}
and in \cite{CKL92}, where it was shown that
the map  (\ref{GenCHOI}) is positive if and only if
\begin{equation}
a\ge 1, \quad a+b+c\ge 3, \quad
 1\le a\le 2 \Longrightarrow bc\ge (2-a)^2,
 \label{Choipos}
\end{equation}
while it is decomposable if and only if
\begin{equation}
a\ge 1, \quad  1\le a\le 3 \Longrightarrow bc\ge (3-a)^2/4.
\label{Choiindec}
\end{equation}
In particular, $\Psi_{2,0,c}$ is positive but not decomposable for $c\ge 1$.
All generalized indecomposable Choi maps are known to be {\sl atomic}
\cite{HaKC98},
it is they cannot be written as a convex sum of $2$--positive and
$2$--co--positive maps \cite{TT88}.
An example of an indecomposable map belonging to ${\cal P}_4$ was given
by Robertson \cite{Ro83}.
A family of indecomposable maps for an arbitrary finite dimension $N\ge 3$
was recently found by Kossakowski \cite{Ko03}.
They consist of an affine contraction of the set  ${\cal M}^{(N)}$
of density  matrices into the ball
inscribed in it followed by a generic
rotation from $O(N^2-1)$.
Although several other methods of construction of indecomposable maps
were proposed  \cite{Tan86,TT88,Os91,KK94}, some of them in the
context of quantum entanglement \cite{Te00,Yu00,HKP03},
the general problem of describing all positive
maps remains open. In particular, it is not known, if one could find a
finite set of $K$ positive maps $\{\Psi_j\}$, such that
${\cal P}_N={\rm conv~~ hull~}\Bigl( \cup_{j=1}^K
\Psi_j( {\cal CP}_N )\Bigr) $.

Due to the theorem of St{\o}rmer and Woronowicz
the answer is known only for $N=2$, for which $K=2$, $\Psi_1={\mathbbm
1}$ and $\Psi_2=T$.  As emphasized by Horodeccy
in an important work \cite{HHH96a}
these properties of the set ${\cal P}_N$ become decisive for the
separability problem: the separability criterion
based on the positivity of $({\mathbbm 1} \otimes T)\rho$
works for the system of two qubits,
but does not solve the problem in the general case of the $N \times N$
composite system.

\medskip

\begin{figure} [htbp]
   \begin{center}
\
 \includegraphics[width=10.0cm,angle=0]{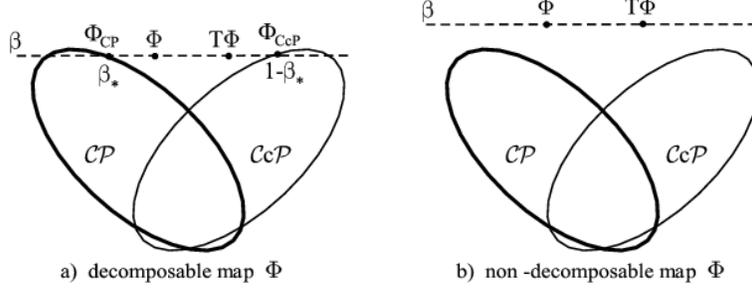}
\caption{Geometric criterion to verify decomposability of a map
$\Phi$: (a) if the line passing through $\Phi$ and $\Phi T$
crosses the set of completely positive maps,
a decomposition of $\Phi$ is explicitly constructed.}
 \label{fig:oper7}
\end{center}
 \end{figure}

The indecomposable maps are worth investigating, since each
such map provides a criterion for separability.
Conditions for a positive map $\Phi$ to be decomposable
were found some time ago by St{\o}rmer \cite{St82}.
Since this criterion is not a constructive one,
we describe here a simple test which may confirm the decomposability.
Assume first that the map is not symmetric with respect to the
transposition $\Phi \ne T\Phi $.
These two different points determine a line in the space of maps,
parametrized by $\beta$,
along which we analyze the dynamical matrix
\begin{equation}
D_{\beta \Phi+(1-\beta)T\Phi}=
   \beta D_{\Phi} +(1-\beta)D_{\Phi}^{T_A}
 \label{decomcheck}
\end{equation}
and check its positivity by diagonalization.
Assume that this matrix is found to be positive for some $\beta_* <0 $
(or $\beta_* >1$), then the line
(\ref{decomcheck})  crosses the set of completely positive
maps (see  Fig. \ref{fig:oper7}a).
Since $D(\beta_*)$ represents a CP map $\Psi_{\rm CP}$,
hence $D(1-\beta_*)$ defines a completely co--positive map
 $\Psi_{\rm CcP}$,
and we find an explicit decomposition,
$\Phi=[-\beta_* \Psi_{\rm CP}+(1-\beta_*)\Psi_{\rm CcP}]/(1-2\beta_*)$.
In this way the decomposability of $\Phi$ may be established,
but with this criterion one
cannot confirm that a given map is indecomposable.

\vskip -0.4cm
\medskip

To study the geometry of the set of positive maps it is convenient to
work with the Hilbert--Schmidt distance, defined by the $HS$ norm of the
difference between the superoperators,
$d(\Psi,\Phi)=d_{HS}(L_{\Psi},L_{\Phi})=
||L_{\Psi}-L_{\Phi}||_{HS}$.
Since the reshuffling of a matrix does not influence its $HS$ norm,
the distance can be measured directly in the space of dynamical matrices,
$d(\Psi,\Phi)=d_{HS}(D_{\Psi},D_{\Phi})$.
Note that for unital one qubit maps, (\ref{densaDtau})
with ${\vec \kappa}=0$,
one has $d(\Phi_1,\Phi_2 )=|{\vec \eta}_1- {\vec \eta}_2 |$,
so Fig. \ref{fig:oper3b}
represents correctly the HS geometry of the space of
$N=2$ unital maps.

In order to characterize, to what extent a given map
$\Psi$ acting on ${\cal M}^{(N)}$
is close to the boundary of the set of positive (CP or CcP) maps,
let us introduce the following quantities

\begin{eqnarray} 
\! \! {\bf a)} & {\it \quad complete \quad positivity, \quad} &
cp(\Phi)\equiv \min_{{\cal M}^{(N)}} \langle \rho |D_{\Phi}|\rho\rangle, 
\label{comppos}
\\
{\bf b)} &{\it \  complete \   co\!-\!positivity, \ } &
ccp(\Phi)\equiv \min_{{\cal M}^{(N)}} \langle \rho
|D^{T_A}_{\Phi}|\rho\rangle \ , \\
\! \! {\bf c)} &
{\it  \quad positivity , \quad} &
p(\Phi)\equiv \! \! \!\! \min_{x,y\in {\mathbb C}P^{N-1}} \! \! \!
[\langle x\otimes y|D_{\Phi}|x
\otimes y\rangle] .
\label{cpos}
\end{eqnarray}

The two first quantities may be easily found by diagonalization,
$cp(\Phi)=\min \{ {\rm eig} (D_{\Phi}) \}$ and
$ccp(\Phi)=\min \{ {\rm eig} (D^{T_A}_{\Phi}) \}$.
Although $p(\Phi)\ge cp(\Phi)$
by construction
 the evaluation of positivity is more involved,
since one needs to perform the minimization over the
space of all product states, i.e. the Cartesian product
${\mathbb C}P^{N-1} \times {\mathbb C}P^{N-1}$.
No straightforward method of computing this minimum is known,
so one has to rely on numerical minimization.
In certain cases this quantity was estimated analytically
by Terhal \cite{Te00}
and numerically by G{\"u}hne et al. \cite{GHDELMS02,GHDELMS03}
in the context of characterizing the entanglement witnesses.
In fact a non-positive dynamical matrix $D$,
which describes a non completely positive map,
 may be just considered as an entanglement witnesses --
an operator $D$ such Tr$D\rho$ is not negative for
all separable states and negative for
a given entangled state \cite{LKCHC00,LKHC01,Te02,Br02,PR03}.

As follows from the positivity of $D$ and $D^{T_A}$ and
the property of block positivity (\ref{dynmatr3b}), a given map $\Phi$
is completely positive (CcP, positive) if and only if the
complete positivity (ccp, positivity) is non--negative.
As marked in Fig. \ref{fig:oper6}b, the relation $cp(\Phi)=0$
defines the boundary of the set ${\cal CP}_N$,
while $ccp(\Phi)=0$ and $p(\Phi)=0$ define the boundaries of
${\cal C}c{\cal P}_N$ and ${\cal P}_N$.
By direct diagonalization of the dynamical matrix we find that
$cp({\mathbbm 1})= ccp(T)=0$ and
$ccp({\mathbbm 1})=cp(T)=-1$.

For any not completely positive map $\Phi_{\rm nCP}$ one may
look for its best approximation
with a physically realizable
$CP$ map $\Phi_{\rm CP}$,
e.g. by minimizing their HS  distance
$d(\Phi_{\rm nCP},\Phi_{\rm CP})$ --
see Fig. \ref{fig:oper8}a.
Such maps, called
{\sl structural physical approximation} were introduced in \cite{HE02}
to propose an  experimentally feasible scheme of entanglement
detection and later studied in \cite{Fi02}.

To see a simple application of complete positivity,
consider a non physical positive map with
$cp (\Phi_{\rm nCP})=-x<0$.
Its possible CP approximation, but generally not the optimal one,
may be constructed out of its convex combination
with the completely depolarizing channel $\Psi_*$.
Diagonalizing the dynamical matrix representing the map
$\Psi_x=a\Phi_{\rm nCP}+(1-a)\Psi_*$
with $a=1/(Nx+1)$
we see that its smallest eigenvalue is equal to zero,
so $\Psi_x$ belongs to the boundary of ${\cal CP}_N$.
Hence the distance $d(\Phi_{\rm nCP},\Phi_x)$,
which is a function of the complete positivity
$cp(\Phi_{\rm nCP})$, gives an upper bound for the
distance of $\Phi_{\rm nCP}$ from the set ${\cal CP}$.
In a similar way one may use $ccP(\Phi)$
to obtain an upper bound for the distance of an analyzed
non CcP map $\Phi_{\rm nCcP}$ from the set ${\cal C}c{\cal P}$.
Interestingly,
the solution of the analogous problem in the space of density matrices
allows one to characterize the entanglement of a two-qubit mixed state
$\rho_1$ by its minimal distance to the set of separable states.
In the two--qubit system the entangled states 
do not have positive partial transpose, so $T_A(\rho_1)$
is not a state.  
As shown in Fig. \ref{fig:oper8}b one may also 
study a related problem of finding the state $\rho_2$ which is 
closest to $T_A(\rho_1)$.

\section{Jamio{\l}kowski isomorphism}
\label{sec:jamiol}

Let ${\cal CP}_N$ denote the space of all
trace preserving, completely  positive maps
$\Phi:{\cal M}^{(N)} \to {\cal M}^{(N)}$.
Note that this is a convex set.
Any such a map  $\Phi$ may be uniquely represented by its dynamical
 matrix $D_{\Phi}$ of size $N^2$. It is a positive,
 Hermitian matrix and its trace is equal to
$N$. Hence the rescaled matrix $\rho_{\Phi}\equiv D_{\Phi}/N$
represents a mixed state in ${\cal M}^{(N^2)}$,
and the entropy of the operation $S(\Phi)$
equals to the von Neumann entropy of $\rho_{\Phi}$.
In fact rescaled dynamical matrices explore only a subspace
 of this set  determined by the trace
preserving conditions
(\ref{dynmatr3c}), which impose $N^2$ constraints.
Let us denote this $N^4-N^2$ dimensional
set by  ${\cal M}_I^{(N^2)}$.
Since for any trace preserving CP map we may find a dynamical matrix,
and vice versa,
the correspondence between the maps from
${\cal CP}_N$ and the states of
${\cal M}_I^{(N^2)}$ is one--to--one.
In Table \ref{tab:jam} this isomorphism is labeled by $J_{III}$.

\bigskip
\begin{table}
\caption{Jamio{\l}kowski Isomorphism (\ref{jamiol2})
between Hermicity reserving linear maps $\Phi$ definded
on the space of mixed states ${\cal M}^{(N)}$ which act on ${\cal H}_N$ and the
operators $D_{\Phi}$ on the composed space
${\cal H}_N \otimes {\cal H}_N$. }
\smallskip
{\renewcommand{\arraystretch}{1.45}
\begin{tabular}
[c]{||c|c|c||}
\hline \hline
Isomorphism &
\parbox {4.4cm}{\centering Linear maps \\
                 $\Phi: {\cal M}^{(N)} \to  {\cal M}^{(N)}$ } &
\parbox {4.4cm}{\centering Hermitian operators \\
                 $D_{\Phi}: {\cal H}_{N^2} \to  {\cal H}_{N^2}$ } \\
\hline
$J_I$ & \parbox {2.5cm}{ positive maps $\Phi$} &
 \parbox {4.4cm}{\centering operators $D$ \\ positive  on product states} \\
\hline
$J_{II}$ & \parbox {4.4cm}{\centering completely  positive maps $\Phi$}
& \parbox {4.4cm}{\centering positive operators $D$ } \\
\hline
$J_{III}$  & \parbox {4.4cm}{\centering quantum operations: \\ CP,
trace preserving maps}
  & \parbox {4.3cm}{\centering states $\rho=D/N$ \  such that
Tr$_A D={\mathbbm 1}$} \\
\hline
$J_{IIIa}$ & \parbox {4.4cm}{\centering completely positive, \ unital maps}
& \parbox {4.3cm}{\centering states $\rho=D/N$ \  such that
Tr$_B D={\mathbbm 1}$} \\
\hline
$J_{IV}$  &
 \parbox {4.4cm}{\centering CP and CcP    \\ trace preserving maps}
  &\parbox {4.3cm}{\centering mixed states $\rho=D/N$ \\    such that
Tr$_A D={\mathbbm 1}$ with  \\
     positive partial transpose }   \\ \hline
\parbox {2.5cm}{\centering example \\ of $J_{IV}$ \\ for $N=2$ }
 & \parbox {4.4cm}{\centering CP and CcP \\ trace preserving maps \\
                     $\Phi: {\cal M}^{(2)} \to  {\cal M}^{(2)}$ }
 & \parbox {4.3cm}{\centering  separable states $\rho=D/N$ \\ of a
      $ 2 \times 2$ composite system
      \\  such that Tr$_A D={\mathbbm 1}$} \\
\hline
$J_{IVa}$ & \parbox {4.4cm}
{\centering unitary  rotations \\
  $\rho' =  U \rho U^{\dagger}$, \\
  $ D_{\Phi}=(U \otimes U^*)^R $}
& \parbox {4.3cm}{\centering maximally entangled \\ pure states \\
$(U \otimes {\mathbbm 1})|\psi\rangle\langle \psi|$ } \\
\hline
\parbox {2.5cm}{\centering $N=2$ example \\
of $J_{IVa}$ }
 & ${\mathbbm 1} \leftrightarrow (1,0, 0, 1)$
 & \parbox {4.4cm}{\centering
   $|\psi_+\rangle\equiv \frac{1}{\sqrt{2}} |00\rangle +|11\rangle$ }    \\
\parbox {2.5cm}{\centering {\sl Pauli matrices} } &
${\sigma_x}\leftrightarrow (0,1,1,0)$
& \parbox {4.4cm}{\centering
   $|\phi_+\rangle \equiv \frac{1}{\sqrt{2}} |01\rangle +|10\rangle$ }    \\
\parbox {2.5cm}{\centering  versus} &
 ${\sigma_y} \leftrightarrow (0,-i, i,0)$
 & \parbox {4.4cm}{\centering
$|\phi_-\rangle \equiv \frac{1}{\sqrt{2}} |01\rangle -|10\rangle$ } \\
\parbox {2.5cm}{\centering {\sl Bell states} \\
 $\rho_{\phi}=| \phi\rangle \langle \phi |$} &
${\sigma_z} \leftrightarrow (1,0, 0, -1)$
 & \parbox {4.4cm}{\centering
   $|\psi_-\rangle \equiv \frac{1}{\sqrt{2}} |00\rangle -|11\rangle$ }    \\
\hline
$J_V$ &  \parbox {4.4cm}{\centering completely depolarizing  \\
  channel $\Phi_*$ }      & \parbox {4.3cm}{\centering
 maximally mixed \\ state  $\rho_*={\mathbbm 1}/N$ }   \\
 \hline \hline
\end{tabular}
}
\label{tab:jam}
\end{table}

Let us find the dynamical matrix for the identity operator, $L={\mathbbm
1}_{N^2}$,
\begin{equation}
L^{\mathbbm 1}_{\stackrel{\scriptstyle m \mu}{n \nu}}=
\delta_{mn}\delta_{\mu\nu}
{\quad \rm so \quad that \quad}
D^{\mathbbm 1}_{\stackrel{\scriptstyle m \mu}{n \nu}} =
(L^{\mathbbm 1}_{\stackrel{\scriptstyle m \mu}{n \nu}})^R =
\delta_{m\mu}\delta_{n \nu}
=N \rho^{\psi}_{\stackrel{\scriptstyle m \mu}{n \nu}},
 \label{idenentan}
\end{equation}
where
$\rho^{\psi}= |\psi\rangle \langle \psi|$
represents the operator of projection on
the maximally entangled state
 of the composite system
\begin{equation}
|\psi\rangle=\sum_{i=1}^N \frac{1}{\sqrt{N}} |i\rangle \otimes
|i\rangle.
 \label{maxenta}
\end{equation}
This state is written in its Schmidt form \cite{Pe95b} 
(for $N=2$ it is the famous Bell state), and we see
that all its Schmidt coefficients are equal,
$\lambda_1=\lambda_i=\lambda_N=1/N$.
Thus we have found that the
identity operator corresponds to the maximally entangled pure state
$|\psi\rangle\langle \psi|$ of the composite system.
Interestingly, this correspondence may be extended for other operations,
or in general, for arbitrary linear maps.
Any linear map $\Phi$ acting on the space of mixed states ${\cal
M}^{(N)}$ can be associated,
 via its dynamical matrix $D_{\Phi}$,  with an
operator acting in the enlarged Hilbert state
  ${\cal H}_N \otimes {\cal H}_N$
\begin{equation}
 \Phi: {\cal M}^{(N)} \to  {\cal M}^{(N)}
\quad \longleftrightarrow \quad D_{\Phi}= N
\bigl[ \Phi \otimes {\mathbbm 1}\bigr] ( |\psi\rangle\langle \psi|)
 \label{jamiol2}
\end{equation}
To show this we represent the linear map $\Phi$
 by its matrix $L$ introduced
in (\ref{dynmatr1}),
write the operator $\Phi \otimes {\mathbbm 1}$
as an eight--indices matrix
 and study its action on the state $\rho^{\psi}$
expressed by (\ref{idenentan}),
\begin{equation}
L_{\stackrel{\scriptstyle m n}{m'n'}}
{\mathbbm 1}_{\stackrel{\scriptstyle \mu \nu}{\mu'\nu'}}
N \rho^{\psi}_{\stackrel{\scriptstyle m' \mu'}{n'\nu'}}
= L_{\stackrel{\scriptstyle m n}{\mu \nu}}=
D_{\stackrel{\scriptstyle m\mu}{n \nu}} .
 \label{jamiol3}
\end{equation}
An analogous
 operation ${\mathbbm 1} \otimes \Phi$ acting on $\rho^{\psi}$
leads to the matrix $D^F$ with the same spectrum.
Conversely, for any positive matrix $D$
we find the corresponding map $\Phi$
by diagonalization.
The reshaped eigenvectors $|\chi_i\rangle$ of $D$,
rescaled by the roots of its eigenvalues give the
canonical Kraus form ({\ref{dynmatr5},\ref{dynmatr6})
of the corresponding operation $\Phi$.
Furthermore, the entropy of
 the quantum operation
$S(\Phi)$ equals the von Neumann entropy $S(\rho_{\Phi})$ 
of the corresponding state $\rho_{\Phi}=D_{\Phi}/N$.

\vskip -0.5cm
\begin{figure} [htbp]
   \begin{center}
\
 \includegraphics[width=10.0cm,angle=0]{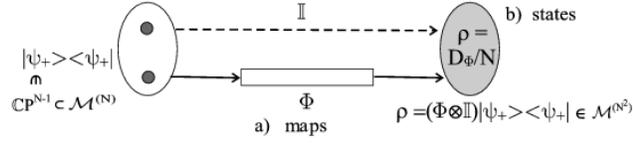}
\caption{Duality between quantum map $\Phi$ acting on a
part of the maximally entangled state $|\psi_+\rangle$
and the resulting density matrix $\rho=\frac{1}{N} D_{\Phi}$}
 \label{fig:oper9}
\end{center}
 \end{figure}
\medskip

Consider now a more general case in which $\rho$ denotes a state acting
on a composite Hilbert space ${\cal H}_N\otimes {\cal H}_N$.
Let $\Phi$ be an arbitrary map which sends ${\cal M}^{(N)}$ into
itself and let $D_{\Phi}=L^R_{\Phi}$ denote its dynamical matrix (of
size
$N^2$). Acting with the extended map on $\rho$ we find its image
$\rho'=[\Phi \otimes {\mathbbm 1}](\rho)$.
Writing down the explicit form of the corresponding
linear map in analogy to (\ref{jamiol3}), and contracting over the four
indices representing $\mathbbm 1$, 
we obtain
\begin{equation}
 (\rho')^R=L_{\Phi}\rho^R
{\quad \rm so \quad that  \quad}
 \rho'=(D_{\Phi}^R \rho^R)^R .
 \label{rhoDR}
\end{equation}
In the above formula the standard
multiplication of square matrices takes place,
in contrast to  Eq. (\ref{dynmatr1}) in which the state
$\rho$ acts on a simple Hilbert space and is treated as a vector.

Note that Eq. (\ref{jamiol2}) may be obtained as a special case of
(\ref{rhoDR}) if one takes for $\rho$ the maximally entangled state
(\ref{maxenta}), for which $(\rho^{\psi})^R={\mathbbm 1}$.
Formula (\ref{rhoDR}) provides a useful application of the
dynamical matrix corresponding to a map $\Phi$ acting on a subsystem.
Since the normalization of matrices does not influence positivity, this
result implies the reshuffling lemma (\ref{reshlem}).

Formula (\ref{jamiol2}) may also be used
to find operators $D$ associated with positive maps $\Phi$
which are neither trace preserving nor complete positive.
The correspondence between the set of positive
linear maps and dynamical
matrices acting in the composite space and positive on product states is
called {\sl Jamio{\l}kowski isomorphism}
since it follows from his results
obtained in \cite{Ja72}.
Some aspects
of the duality between maps and states were recently investigated in
\cite{Ha03,AP03b}.
Let us mention here explicitly certain special
cases of this isomorphism labeled by $J_I$ in Table \ref{tab:jam}.
The set of all completely positive maps $\Phi$ is isomorphic to the set
of all positive matrices $D$, (case $J_{II}$).
Unital CP maps are related to dynamical matrices
which satisfy ${\rm tr}_B D = {\mathbbm 1}$ (case $J_{IIIa}$).
The set of all quantum operation,
(i.e. the trace preserving, CP maps)
corresponds to the set of positive matrices $D$
fulfilling another constraint,
${\rm tr}_A D = {\mathbbm 1}$
-- see  Table \ref{tab:jam}, item $J_{III}$.
An apparent asymmetry between the role of both subsystems is
due to the particular choice of the
relation (\ref{jamiol2}); if
the operator ${\mathbbm 1}\otimes \Phi $ is used instead,
the subsystems $A$ and $B$ in the partial trace constraints
need to be interchanged.

An important case $J_{IV}$  of the isomorphism
concerns the states with {\sl positive partial transpose},
$\rho^{T_A}\ge 0$. 
called briefly PPT states. 
Another case, $J_{IVa}$, relates  
the set of unitary rotations,
 $\rho'=\Phi(\rho)=U\rho U^{\dagger}$ 
with the maximally entangled states,
$(U \otimes {\mathbbm 1}) \bigr |\psi\rangle\langle\psi|$.
The local unitary operation
$(U \otimes {\mathbbm 1})$
preserves the purity of a state and its Schmidt coefficients.
Thus the set of  unitary matrices $U$ of size $N$
is isomorphic to the set of the maximally entangled pure states
of the composite  $N \times N$ system.
In particular,
vectors obtained by reshaping the Pauli matrices $\sigma_i$
represent the Bell states in the computational basis,
as listed in Table \ref{tab:jam}.
Eventually,  case $J_{V}$ consists of a single, distinguished point
in both spaces: the completely depolarizing channel $\Phi_*$
and the corresponding maximally mixed state $\rho_*$.
Note the following inclusion relations
of the sets mentioned in Table \ref{tab:jam},
$J_I \supset J_{II} \supset J_{III} \supset J_{IV} \supset J_V
\subset J_{IIIa}$
and $J_{II} \supset J_{IIIa} \supset J_{IVa} \subset J_{III}$,
as sketched in Fig. \ref{fig:oper8}.

\medskip

\section{Quantum maps and quantum states}
\label{sec:duality}

Relation (\ref{jamiol2})
allows one to link an arbitrary linear map $\Phi$ with
the corresponding linear operators given by the dynamical matrix
$D_{\Phi}$. Expressing the maximally entangled
state  $|\psi\rangle$ in (\ref{jamiol2})
by its Schmidt form  (\ref{maxenta})
we may compute the matrix elements of
$D_{\Phi}$ in the product basis consisting of the states
$|i\otimes j\rangle$.
Due to the factorization of the right hand side
we see that the double sum describing $\rho_{\Phi}=D_{\Phi}/N$
drops out and the result reads
\begin{equation}
\langle k\otimes i | D_{\Phi}| l \otimes j \rangle =
\langle k \bigl| \Phi(| i\rangle \langle j|) \bigr| l \rangle .
 \label{jamiobis}
\end{equation}
This equation may also be understood as a definition of a
map $\Phi$ related to the linear operator $D_{\Phi}$.
It proves the isomorphism $J_I$ from Table \ref{tab:jam}:
if $D_{\Phi}$ is block positive,
then the corresponding map $\Phi$ sends positive operators into
positive operators \cite{Ja72}.

As listed in Table \ref{tab:jam}
and shown in Fig. \ref{fig:oper8}
the Jamio{\l}kowski isomorphism (\ref{jamiobis})
may be applied in various setups \cite{AP03b}.
Relating linear maps from ${\cal P}_N$
with the operators acting on an extended space
${\cal H}_{N} \otimes {\cal H}_{N}$
we may compare:

{\bf i)} individual objects, e.g. completely depolarizing channel
$\Phi_*$ and the maximally mixed state $\rho_*$,

{\bf ii)} families of objects, e.g. the depolarizing
channels and the Werner states \cite{We89},

{\bf iii)} entire sets, e.g. the set of
${\cal CP} \cup {\cal C}c{\cal P}$ maps and the set of PPT states,

{\bf iv)} certain problems, e.g. for an arbitrary CP map $\Phi$ find the
closest CcP map, and the problem of finding the PPT state
(separable state for $N=2$) closest to an arbitrary (entangled) state, \ and

{\bf v)} their solutions...

To get some more experience concerning the analyzed duality 
between quantum maps and quantum states 
compare both sides of Fig. \ref{fig:oper8}.
 Note that this illustration
may also be considered as a strict representation of a fragment of
the space of one--qubit unital maps (a) or the space of two-qubits
density matrices in the HS geometry (b).
It is nothing else but the cross--section of the cube
representing the positive maps in Fig. \ref{fig:oper3b}c
along the plane determined  by ${\mathbbm 1},T$ and $\Phi_*$.

\vskip -0.3cm
\begin{figure} [htbp]
   \begin{center}
\
 \includegraphics[width=13.0cm,angle=0]{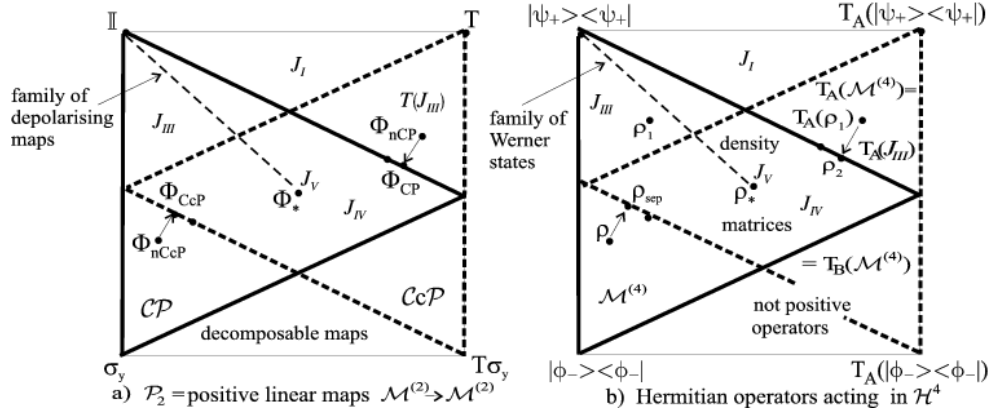}
\caption{Isomorphism between objects, sets, and problems:
a) linear one qubit maps, b) linear operators
acting in two-qubit Hilbert space ${\cal H}_4$.
Labels $J_i$ refer to the sets related by the isomorphisms 
defined in Table \ref{tab:jam}.}
 \label{fig:oper8}
\end{center}
 \end{figure}
\medskip

Let us finish this section by pointing out
an analogue of the Jamio{\l}kowski isomorphism
in the classical case.
The space of all classical states --  probability
vectors of size $N$ -- forms the $N-1$ dimensional simplex $\Delta_{N-1}$.
A discrete dynamics in this space is given by a stochastic
transition matrix
$T_N:\Delta_{N-1} \to \Delta_{N-1}$. It contains non--negative entries,
and due to stochasticity
the sum of all its elements is equal to $N$.
Hence reshaping the transition matrix $T_N$
and rescaling it by $1/N$ we receive a probability vector ${\vec t}$ of
length $N^2$. The classical states defined in this way
form a measure zero, $N(N-1)$ dimensional, convex subset of
$\Delta_{N^2-1}$.
Consider, for instance, the set of $N=2$ stochastic matrices,
which can be parametrized as
{\small
 $T_2= \left[ \begin{array}
[c]{cc}
a & b \\
1-a & 1-b
\end{array}  \right]$
}
with $a,b\in [0,1]$. The set of the corresponding
probability vectors ${\vec t}=(a,b,1-a,1-b)/2$
forms a square of size $1/2$ -- the maximal square which may be
inscribed into the unit
tetrahedron $\Delta_3$ of all $N=4$ probability vectors.

The classical dynamics may be considered as a (very) special
subclass of quantum dynamics, defined on the set of diagonal density
matrices. Hence the classical and quantum duality between maps and
states may be succinctly summarized in the following, commutative
diagram
\begin{equation}
\begin{array}{llcl}
{\rm quantum: \quad \quad} &
\bigr[ \Phi :{\cal M}^{(N)}\rightarrow   {\cal M}^{(N)}\bigl] &
\longrightarrow  & D_{\Phi} \in {\cal M}^{(N^{2})} \\
&
\quad \quad {\vphantom{\Bigg|}\downarrow }  {\quad \rm maps} &
 & {\vphantom{\Bigg|}\downarrow }
{\quad \rm states}\\
{\rm classical: \quad \quad} &
\bigr[ T:\Delta _{N-1}\rightarrow \Delta _{N-1}\bigl] &
 \longrightarrow & \ {\vec t} \ \in \Delta _{N^{2}-1}  \ .
\end{array}
\label{diagr}
\end{equation}
Alternatively, vertical arrows may be interpreted as
the action of the coarse graining operation $\Psi_{\rm CG}$
defined in section \ref{sec:bistoch}.
For instance, for the trivial (do nothing)
one--qubit quantum map $\Phi_{\mathbbm 1}$, the superoperator
$L={\mathbbm 1}_4$ restricted to diagonal matrices
gives the identity matrix, $T={\mathbbm 1}_2$,
and the classical state ${\vec t}=(1,0,0,1)/2\in \Delta_3$.
On the other hand, this very vector represents
the diagonal of the maximally entangled state
$\frac{1}{2}D_{\Phi}=|\psi\rangle\langle \psi|$.
To prove the commutativity of the diagram (\ref{diagr})
in the general case define
the stochastic matrix $T$
as a submatrix of the superoperator (\ref{dynmatr1}),
$T_{mn}=L_{\stackrel{\scriptstyle mm}{nn}}$ (left vertical arrow).
Note that the vector $\vec t$ obtained by its reshaping
satisfies ${\vec t}=
{\rm diag}(L^R)={\rm diag}(D_{\Phi})$, so
it represents the diagonal of the dynamical matrix (right arrow).

\section{ Envoi}

\vspace{5mm}

In this work we analyzed the space ${\cal P}_N$ of positive maps,
which send the set of mixed quantum states ${\cal M}^{(N)}$ into itself,
and its various subsets: the set  ${\cal CP}_N$ of completely positive maps,
and the set ${\cal B}_N$ of bistochastic maps. 
For any quantum map $\Psi$ we have introduced three
quantities,  (\ref{comppos}-\ref{cpos}),
to characterize the location of $\Psi$ with respect 
to the boundaries of the sets
of positive (completely positive or completely co-positive) maps.
 We have defined and investigated the set of unistochastic maps, the elements
of which are
determined by a unitary matrix $U$ of size $N^2$ and correspond to
the coupling with the $N$-dimensional environment, initially in 
the maximally mixed state.

The spaces of quantum maps (stochastic, bistochastic, unitary evolutions),
which act on the set ${\cal M}^{(N)}$ of mixed quantum states,
may be related with the corresponding classical maps
(stochastic, bistochastic, permutations), which act 
on the simplex $\Delta_{N-1}$ of all $N$--point probability distributions.
We have extended the Jamio{\l}kowski isomorphism, linking the space of
quantum maps with the space of bipartite quantum states, 
to the classical case (\ref{diagr}).

In general the state--map  duality, analyzed
in this paper, may be formulated and applied in
a variety of contexts and set ups.
For instance, the set of $SU(4)$ matrices 
may be considered as:

a) the space of maximally entangled states of a composite,
$4\times 4$ system, $|\psi\rangle \in  {\mathbb C}P^{15}\subset
{\cal M}^{(16)}$,

b) the set of two-qubit unitary
gates  acting on ${\cal M}^{(4)}$ -
see  e.g.
\cite{KC01,HVC02,ZVWS03,NDDGMOBHH02}, 

c) the set ${\cal BU}_2$ of one qubit unistochastic operations
 (\ref{unitevol5}),
          $\Psi_U \in {\cal BU}_2\subset {\cal B}_2\subset {\cal CP}_2$.

To conclude, we would like to convey two main points. 
On one hand, the set of positive maps has an interesting geometry, which is
worthy of investigation in the general case $N\ge 3$.
On the other hand, the marvelous  duality between quantum maps 
and quantum states, based on the Jamio{\l}kowski isomorphism,
allows one to apply all the knowledge gained in studying 
the set of quantum maps to describe the set of quantum states  
(or conversely). For instance, the known structure of
the set of completely positive maps relates to the structure of 
all quantum states, while the description of the (larger) set
of all positive maps would allow us to describe the subset of all separable
states.

It is a pleasure to thank  R.~Alicki, R.~Dobrza\-{\'n}ski--Demko\-wicz, 
 T.~Havel,
P.~Horodecki, R.~Horodecki, A.~Kossakowski,
M.~Ku{\'s}, W.~A.~Majewski and F.~Mintert
for fruitful discussions and helpful comments.
Financial support by Komitet Bada{\'n} Naukowych
under the grant PBZ-Min-008/P03/03
and Swedish grant from VR is gratefully acknowledged.

\bibliographystyle{abbrv}
\bibliography{zbosid}

\begin{thebibliography}{100}

\bibitem{AA03}
J.~{\AA}berg.
\newblock Subspace preserving completely positive maps.
\newblock {\em preprint quant-ph}, page 0302180, 2003.

\bibitem{Ac76}
L.~Accardi.
\newblock Nonrelativistic quantum mechanic as a non\-commu\-ta\-tive {M}arkof
  process.
\newblock {\em Adv. Math.}, 20:329, 1976.

\bibitem{ABHH+01}
G.~Alber, T.~Beth, M.~Horodecki, P.~Horodecki, R.~Horodecki, M.~R{\"o}tteler,
  H.~Weinfurter, R.~Werner, and A.~Zeilinger.
\newblock {\em Quantum Information: An Introduction to Basic Theoretical
  Concepts and Experiments}.
\newblock Springer, Berlin, 2001.
\newblock Springer Tracts in Modern Physics.

\bibitem{ACF03}
S.~Albeverio, K.~Chen, and S.-M. Fei.
\newblock Generalized reduction criterion for separability of quantum states.
\newblock {\em Phys. Rev.}, A 68:062313, 2003.

\bibitem{AF01}
R.~Alicki and M.~Fannes.
\newblock {\em Quantum Dynamical Systems}.
\newblock Oxford University Press, Oxford, 2001.

\bibitem{AL87}
R.~Alicki and K.~Lendi.
\newblock {\em Quantum Dynamical Semigroups and Applications}.
\newblock Springer--Verlag, Berlin, 1987.

\bibitem{AHW00}
G.~G. Amosov, A.~S. Holevo, and R.~F. Werner.
\newblock On some additivity problems in quantum information theory.
\newblock preprint math-ph/0003002, 2000.

\bibitem{AP03b}
P.~Arrighi and C.~Patricot.
\newblock On quantum operations as quantum states.
\newblock {\em preprint quant-ph}, /:0307024, 2003.

\bibitem{Ar69}
W.~B. Arveson.
\newblock Subalgebras of $c^*$--algebras.
\newblock {\em Acta Math.}, 123:141, 1969.

\bibitem{AMM97}
Arvind, K.~S. Mallesh, and N.~Mukunda.
\newblock A generalized \uppercase{P}ancharatnam geometric phase formula for
  three-level quantum systems.
\newblock {\em J. Phys.}, A 30:2417, 1997.

\bibitem{BS93}
C.~Beck and F.~Schl{\"o}gl.
\newblock {\em Thermodynamics of Chaotic Systems}.
\newblock Cambridge University Press, Cambridge, 1993.

\bibitem{BEZ00}
D.~Bouwmeester, A.~K. Ekert, and A.~Z. (eds.).
\newblock {\em The Physics of Quantum Information: Quantum cryptography,
  Quantum Teleportation, Quantum Computation}.
\newblock Springer, Berlin, 2000.

\bibitem{BP02}
H.-P. Breuer and F.~Petruccione.
\newblock {\em The Theory of Open Quantum Systems}.
\newblock Oxford University Press, Oxford, 2002.

\bibitem{Br02}
D.~Bru{\ss}.
\newblock Characterizing entanglement.
\newblock {\em J. Math. Phys.}, 43:4237, 2002.

\bibitem{BK03}
M.~S. Byrd and N.~Khaneja.
\newblock Characterization of the positivity of the density matrix in terms of
  the coherence vector representation.
\newblock {\em quant-ph}, page 03002024, 2003.

\bibitem{CW03}
K.~Chen and L.-A. Wu.
\newblock A matrix realignment method for recognizing entanglement.
\newblock {\em Quant. Inf. Comp.}, 3:193, 2003.

\bibitem{CKL92}
S.-J. Cho, S.-H. Kye, and S.~G. Lee.
\newblock Generalized {C}hoi map in $3$-dimensional matrix algebra.
\newblock {\em Linear Alg. Appl.}, 171:213, 1992.

\bibitem{Cho72}
M.-D. Choi.
\newblock Positive linear maps on $c^{*}$--algebras.
\newblock {\em Can. J. Math.}, 3:520, 1972.

\bibitem{Cho75a}
M.-D. Choi.
\newblock Completely positive linear maps on complex matrices.
\newblock {\em Linear Alg. Appl.}, 10:285, 1975.

\bibitem{Cho75}
M.-D. Choi.
\newblock Positive semidefinite biquadratic forms.
\newblock {\em Linear Alg. Appl.}, 12:95, 1975.

\bibitem{Cho80}
M.-D. Choi.
\newblock Some assorted inequalities for positive maps on $c^*$-algebras.
\newblock {\em J. Operator Theory}, 4:271, 1980.

\bibitem{CL77}
M.-D. Choi and T.~Lam.
\newblock Extremal positive semidefinite forms.
\newblock {\em Math. Ann.}, 231:1, 1977.

\bibitem{Da76}
E.~B. Davies.
\newblock {\em Quantum Theory of Open Systems}.
\newblock Academic Press, London, 1976.

\bibitem{DPS03}
A.~C. Doherty, P.~A. Parillo, and F.~M. Spedalieri.
\newblock A complete family of separability criteria.
\newblock {\em quant-ph}, 1:0308032, 2003.

\bibitem{EK00}
M.-H. Eom and S.-H. Kye.
\newblock Duality for positive linear maps in matrix algebras.
\newblock {\em Math. Scand.}, 86:130, 2000.

\bibitem{Ev84}
D.~E. Evans.
\newblock Quantum dynamical semigroups.
\newblock {\em Acta Appl. Math.}, 2:333, 1984.

\bibitem{EL77}
D.~E. Evans and J.~T. Levis.
\newblock Dilatations of irreversible evolutions in algebraic quantum theories.
\newblock {\em Comm. Dublin. Inst. Adv. Studies}, A 24:., 1977.

\bibitem{Fi02}
J.~Fiur{\'a}{\v s}ek.
\newblock Structural physical approximations of unphysical maps and generalized
  quantum measurements.
\newblock {\em Phys.~Rev.}, A 66:052315, 2002.

\bibitem{FA99}
A.~Fujiwara and P.~Algoet.
\newblock One--to--one parametrization of quantum channels.
\newblock {\em Phys. Rev.}, A 59:3290, 1999.

\bibitem{Gr99}
J.~Gruska.
\newblock {\em Quantum conputing}.
\newblock McGraw--Hill, New York, 1999.

\bibitem{GHDELMS02}
O.~G{\"u}hne, P.~Hyllus, D.~Bru{\ss}, A.~Ekert, M.~Lewenstein, C.~Macchiavello,
  and A.~Sanpera.
\newblock Detection of entanglement with few local measurements.
\newblock {\em Phys. Rev.}, A 66:062305, 2002.

\bibitem{GHDELMS03}
O.~G{\"u}hne, P.~Hyllus, D.~Bru{\ss}, A.~Ekert, M.~Lewenstein, C.~Macchiavello,
  and A.~Sanpera.
\newblock Experimental detection of entanglement via witness operators and
  local measurements.
\newblock {\em J. Mod. Opt.}, 50:1079, 2003.

\bibitem{HaKC98}
K.-C. Ha.
\newblock Atomic positive linear maps in matrix algebra.
\newblock {\em Publ. RIMS Kyoto Univ.}, 34:591, 1998.

\bibitem{HKP03}
K.-C. Ha, S.-H. Kye, and Y.-S. Park.
\newblock Entangled states with positive partial transposes arising from
  indecomposable positive linear maps.
\newblock {\em preprint quant-ph}, page 0305005, 2003.

\bibitem{HVC02}
K.~Hammerer, G.~Vidal, and J.~I. Cirac.
\newblock Characterization of non--local gates.
\newblock {\em Phys. Rev.}, A 66:062321, 2002.

\bibitem{Ha03}
T.~F. Havel.
\newblock Robust procedures for converting among {L}indblad, {K}raus and matrix
  representations of quantum dynamical semigroups.
\newblock {\em J. Math. Phys.}, 44:534, 2003.

\bibitem{HHH96a}
M.~Horodecki, P.~Horodecki, and R.~Horodecki.
\newblock Separability of mixed states: necessary and sufficient conditions.
\newblock {\em Phys. Lett.}, A 223:1, 1996.

\bibitem{HHH01}
M.~Horodecki, P.~Horodecki, and R.~Horodecki.
\newblock Separability of n-particle mixed states: necessary and sufficient
  conditions in terms of linear maps.
\newblock {\em Phys. Lett.}, A 283:1, 2001.

\bibitem{HHH02}
M.~Horodecki, P.~Horodecki, and R.~Horodecki.
\newblock Separability of mixed quantum states: linear contractions approach.
\newblock {\em quant-ph}, page 0206008, 2002.

\bibitem{HHO03}
M.~Horodecki, P.~Horodecki, and J.~Oppenheim.
\newblock Reversible transformations from pure to mixed states and the unique
  measure of information.
\newblock {\em Phys. Rev.}, A 67:062104, 2003.

\bibitem{Ho01}
P.~Horodecki.
\newblock From entanglement witnesses to positive maps: towards optimal
  characterisation of separability.
\newblock In T.~Gonis and P.~E.~A. Turchi, editors, {\em Decoherence and its
  Implications in Quantum Computation and Quantrum Computing}. IOS Press, 2001.

\bibitem{HE02}
P.~Horodecki and A.~Ekert.
\newblock Method for direct detection of quantum entanglement.
\newblock {\em Phys. Rev. Lett.}, 89:127902, 2002.

\bibitem{In75}
R.~Ingarden.
\newblock Quantum information theory.
\newblock {\em Rep. Math. Phys.}, 10:43, 1975.

\bibitem{IK68}
R.~Ingarden and A.~Kossakowski.
\newblock An axiomatic definition of information in quantum mechanics.
\newblock {\em Bull. Acad. Polon. Sci. S{\'e}r. math. astr. phys.}, 16:61,
  1968.

\bibitem{IKO97}
R.~Ingarden, A.~Kossakowski, and M.~Ohya.
\newblock {\em Information Dynamics and Open Systems}.
\newblock Kluver, Dordrecht, 1997.

\bibitem{IU62}
R.~Ingarden and K.~Urbanik.
\newblock Quantum informational thermodynamics.
\newblock {\em Acta Phys. Polon.}, 21:281, 1962.

\bibitem{JS01}
L.~Jak{\'o}bczyk and M.~Siennicki.
\newblock Geometry of {B}loch vectors in two--qubit system.
\newblock {\em Phys. Lett.}, A 286:383, 2001.

\bibitem{Ja72}
A.~Jamio{\l}kowski.
\newblock Linear transformations which preserve trace and positive
  semidefiniteness of operators.
\newblock {\em Rep. Math. Phys.}, 3:275, 1972.

\bibitem{Ja74}
A.~Jamio{\l}kowski.
\newblock An effective method of investigation of positive maps on the set of
  positive definite operators.
\newblock {\em Rep. Math. Phys.}, 5:415, 1975.

\bibitem{Ke02}
M.~Keyl.
\newblock Fundamentals of quantum information theory.
\newblock {\em Phys. Rep.}, 369:431, 2002.

\bibitem{KK94}
H.-J. Kim and S.-H. Kye.
\newblock Indecomposable extreme positive linear maps in matrix algebras.
\newblock {\em Bull. London. Math. Soc.}, 26:575, 1994.

\bibitem{Ki03}
G.~Kimura.
\newblock The {B}loch vector for $n$--level systems.
\newblock {\em Phys. Lett.}, A 314:339, 2003.

\bibitem{KR01}
C.~King and M.~B. Ruskai.
\newblock Minimal entropy of states emerging from noisy channels.
\newblock {\em IEEE Trans. Inf. Th.}, 47:192, 2001.

\bibitem{KL98}
E.~Knill and R.~Laflamme.
\newblock Power of one bit of quantum information.
\newblock {\em Phys. Rev. Lett.}, 81:5672, 1998.

\bibitem{Ko69}
A.~Kossakowski.
\newblock On the quantum informational thermodynamics.
\newblock {\em Bull. Acad. Polon. Sci. S{\'e}r. math. astr. phys.}, 17:349,
  1969.

\bibitem{Ko00}
A.~Kossakowski.
\newblock Remarks on positive maps of finite dimensional simple {J}ordan
  algebras.
\newblock {\em Rep. Math. Phys.}, 46:393, 2000.

\bibitem{Ko03}
A.~Kossakowski.
\newblock A class of linear positive maps in matrix algebras.
\newblock {\em Open Sys. \& Information Dyn.}, 10:1, 2003.

\bibitem{KC01}
B.~Kraus and J.~I. Cirac.
\newblock Optimal creation of entanglement using a two-qubit gate.
\newblock {\em Phys.~Rev.}, A 63:062309, 2001.

\bibitem{Kr71}
K.~Kraus.
\newblock General state changes in quantum theory.
\newblock {\em Ann. Phys.}, 64:311, 1971.

\bibitem{Kr83}
K.~Kraus.
\newblock {\em States, Effects and Operations: Fundamental Notions of Quantum
  Theory}.
\newblock Springer-Verlag, Berlin, 1983.

\bibitem{Ky96}
S.-H. Kye.
\newblock Facial structures for positive linear maps between matrix algebras.
\newblock {\em Canad. Math. Bull.}, 39:74, 1996.

\bibitem{Ky03}
S.-H. Kye.
\newblock Facial structures for unital positive linear maps in the two
  dimensional matrix algebra.
\newblock {\em Linear Alg. Appl.}, 362:57, 2003.

\bibitem{LMM03}
L.~E. Labuschagne, W.~A. Majewski, and M.~Marciniak.
\newblock On k-decomposability of positive maps.
\newblock {\em math-ph}, .:0306017, 2003.

\bibitem{LS93}
L.~J. Landau and R.~F. Streater.
\newblock On {B}irkhoff theorem for doubly stochastic completely positive maps
  of matrix algebras.
\newblock {\em Linear Algebra Appl.}, 193:107, 1993.

\bibitem{LKCHC00}
M.~Lewenstein, B.~Kraus, J.~I. Cirac, and P.~Horodecki.
\newblock Optimization of entanglement witnesses.
\newblock {\em Phys.~Rev.}, A 62:052310, 2000.

\bibitem{LKHC01}
M.~Lewenstein, B.~Kraus, P.~Horodecki, and J.~I. Cirac.
\newblock Characterization of separable states and entanglement witnesses.
\newblock {\em Phys.~Rev.}, A 63:044304, 2001.

\bibitem{Li75}
G.~Lindblad.
\newblock Completely positive maps and entropy inequalities.
\newblock {\em Commun.~Math.~Phys.}, 40:147, 1975.

\bibitem{Li76}
G.~Lindblad.
\newblock On the generators of quantum dynamical semigroups.
\newblock {\em Commun.~Math.~Phys.}, 48:119, 1976.

\bibitem{MaW95}
G.~Mahler and V.~A. Weberrus{\ss}.
\newblock {\em Quantum Networks}.
\newblock Springer, Berlin, 1995, (II ed.) 1998.

\bibitem{Ma75}
W.~A. Majewski.
\newblock Transformations between quantum states.
\newblock {\em Rep. Math. Phys.}, 8:295, 1975.

\bibitem{MM01}
W.~A. Majewski and M.~Marciniak.
\newblock On characterization of positive maps.
\newblock {\em J. Phys.}, A 34:5863, 2001.

\bibitem{MO79}
A.~W. Marshall and I.~Olkin.
\newblock {\em Inequalities: Theory of Majorization and its Applications}.
\newblock Academic Press, New York, 1979.

\bibitem{MKZ04}
M.~Musz, M.~Ku{\'s}, and K.~{\.Z}yczkowski.
\newblock Unitary quantum gates: a measure theoretic approach.
\newblock to be published, 2004.

\bibitem{NC00}
M.~A. Nielsen and I.~L. Chuang.
\newblock {\em Quantum Computation and Quantum Information}.
\newblock Cambridge University Press, Cambridge, 2000.

\bibitem{NDDGMOBHH02}
M.~A. Nielsen, C.~Dawson, J.~Dodd, A.~Gilchrist, D.~Mortimer, T.~Osborne,
  M.~Bremner, A.~Harrow, and A.~Hines.
\newblock Quantum dynamics as physical resource.
\newblock {\em Phys. Rev.}, A 67:052301, 2003.

\bibitem{OP93}
M.~Ohya and D.~Petz.
\newblock {\em Quantum Entropy and Its Use}.
\newblock Springer, Berlin, 1993.

\bibitem{Oi01}
D.~K.~L. Oi.
\newblock The geometry of single qubit maps.
\newblock preprint quant-ph/0106035, 2001.

\bibitem{Os91}
H.~Osaka.
\newblock Indecomposable positive maps in low dimensional matrix algebra.
\newblock {\em Linear Alg. Appl.}, 153:73, 1991.

\bibitem{Os92}
H.~Osaka.
\newblock A class of extremal positive maps in $3 \times 3$ matrix algebras.
\newblock {\em Publ. RIMS. Kyoto Univ.}, 28:747, 1992.

\bibitem{OH85}
C.~J. Oxenrider and R.~D. Hill.
\newblock On the matrix reordering ${\Gamma}$ and ${\Psi}$.
\newblock {\em Linear Alg. Appl.}, 69:205, 1985.

\bibitem{Pe95b}
A.~Peres.
\newblock {\em Quantum Theory: Concepts and Methods}.
\newblock Kluver, Dodrecht, 1995.

\bibitem{Per96}
A.~Peres.
\newblock Separability criterion for density matrices.
\newblock {\em Phys. Rev. Lett.}, 77:1413, 1996.

\bibitem{PR03}
A.~O. Pittenger and M.~H. Rubin.
\newblock Geometry of entanglement witnesses and local detection of
  entanglement.
\newblock {\em Phys. Rev.}, A 67:012327, 2003.

\bibitem{PH81}
J.~A. Poluikis and R.~D. Hill.
\newblock Completely positive and hermitian--preserving transformations.
\newblock {\em Linear Algebra Appl.}, 35:1, 1981.

\bibitem{PBLO03}
D.~Poulin, R.~{Blume--Kohout}, R.~Laflamme, and H.~Olivier.
\newblock Exponential speed--up with a single bit of quantum information.
\newblock {\em quanty-ph}, page 10038, 2003.

\bibitem{Ro83}
A.~G. Robertson.
\newblock Automorphisms of spin factors and the decomposition of positive maps.
\newblock {\em Quart. J. Math. Oxford}, 34:87, 1983.

\bibitem{Rud02}
O.~Rudolph.
\newblock Further results on the cross norm criterion for separability.
\newblock preprint quant-ph/0202121, 2002.

\bibitem{Rud03}
O.~Rudolph.
\newblock Some properties of the computable cross norm criterion for
  separability.
\newblock {\em Physical Review}, A 67:032312, 2003.

\bibitem{RSW02}
M.~B. Ruskai, S.~Szarek, and E.~Werner.
\newblock An analysis of {c}ompletely--{p}ositive {t}race--{p}reserving maps on
  $2 \times 2$ matrices.
\newblock {\em Linear Algebra Appl.}, 347:159, 2002.

\bibitem{St55}
W.~F. Stinespring.
\newblock Positive functions on $c^*$ algebras.
\newblock {\em Proc. Am. Math. Soc.}, 6:211, 1955.

\bibitem{St63}
E.~St{\o}rmer.
\newblock Positive linear maps of operator algebras.
\newblock {\em Acta Math.}, 110:233, 1963.

\bibitem{St82}
E.~St{\o}rmer.
\newblock Decomposable positive maps on $c^*$-algebras.
\newblock {\em Proc. Amer. Math. Soc.}, 86:402, 1982.

\bibitem{St02}
J.~Stry{\l}a.
\newblock Stochastic quantum dynamics.
\newblock arXive preprint quant-ph/0204161, 2002.

\bibitem{SMR61}
E.~C.~G. Sudarshan, P.~M. Mathews, and J.~Rau.
\newblock Stochastic dynamics of quantum--mechanical systems.
\newblock {\em Phys. Rev.}, 121:920, 1961.

\bibitem{SS03}
E.~C.~G. Sudarshan and A.~Shaji.
\newblock Structure and parametrization of stochastic maps of density matrices.
\newblock {\em J. Phys.}, A 36:5073, 2003.

\bibitem{TT83}
T.~Takesaki and J.~Tomiyama.
\newblock On the geometry of positive maps in matrix algebras.
\newblock {\em Math. Z.}, 184:101, 1983.

\bibitem{TT88}
K.~Tanahashi and J.~Tomiyama.
\newblock Indecomposable positive maps in matrix algebra.
\newblock {\em Canad. Math. Bull.}, 31:308, 1988.

\bibitem{Tan86}
W.~Tang.
\newblock On positive linear maps between matrix algebra.
\newblock {\em Linear Alg. Appl.}, 79:33, 1986.

\bibitem{TCDGS99}
B.~Terhal, I.~Chuang, D.~DiVincenzo, M.~Grassl, and J.~Smolin.
\newblock Simulating quantum operations with mixed environments.
\newblock {\em Phys. Rev.}, A 60:88, 1999.

\bibitem{Te00}
B.~M. Terhal.
\newblock A family of indecomposable positive linear maps based on entangled
  quantum states.
\newblock {\em Lin. Alg. Appl.}, 323:61, 2000.

\bibitem{Te02}
B.~M. Terhal.
\newblock Detecting quantum entanglement.
\newblock {\em Theor. Comput. Sci.}, 287:313, 2002.

\bibitem{TV00}
B.~M. Terhal and D.~P. DiVincenzo.
\newblock On the problem of equilibration and the computation of correlation
  functions on a quantum computer.
\newblock {\em Phys. Rev.}, A 61:22301, 2000.

\bibitem{Uh01}
A.~Uhlmann.
\newblock On {$1$}-qubit channels.
\newblock {\em J. Phys.}, A 34:7047, 2001.

\bibitem{VV02}
F.~Verstraete and H.~Verschelde.
\newblock On one-qubit channels.
\newblock preprint quant-ph/0202124, 2002.

\bibitem{We89}
R.~F. Werner.
\newblock Quantum states with {E}instein--{P}odolski--{R}osen correlations
  admitting a hidden--variable model.
\newblock {\em Phys. Rev.}, A 40:4277, 1989.

\bibitem{Wo01}
K.~W{\'o}dkiewicz.
\newblock Stochastic decoherence of qubits.
\newblock {\em Optics Express}, 8:145, 2001.

\bibitem{Wo76a}
S.~L. Woronowicz.
\newblock Nonextendible positive maps.
\newblock {\em Commun. Math. Phys.}, 51:243, 1976.

\bibitem{Wo76b}
S.~L. Woronowicz.
\newblock Positive maps of low dimensional matrix algebra.
\newblock {\em Rep. Math. Phys.}, 10:165, 1976.

\bibitem{YH00}
D.~A. Yopp and R.~D. Hill.
\newblock On completely copositive and decomposable linear transformations.
\newblock {\em Linear Alg. Appl.}, 312:1, 2000.

\bibitem{Yu00}
S.~Yu.
\newblock Positive maps which are not completely entangled.
\newblock preprint quant-ph/0001053, 2000.

\bibitem{ZR02}
C.~Zalka and E.~Rieffel.
\newblock Quantum operations that cannot be implemented using a small mixed
  environment.
\newblock {\em J. Math. Phys.}, 43:4376, 2002.

\bibitem{Za01}
P.~Zanardi.
\newblock Entanglement of quantum evolution.
\newblock {\em Phys. Rev.}, A 63:040304(R), 2001.

\bibitem{ZVWS03}
J.~Zhang, J.~Vala, K.~Whaley, and S.~Sastry.
\newblock Geometric theory of non-local two-qubit operations.
\newblock {\em Phys. Rev.}, A 67:042313, 2003.

\end{thebibliography}

\end{document}